\documentclass[aps,prb,superscriptaddress,twocolumn,showpacs]{revtex4-2}
\usepackage{amsmath}
\usepackage{amssymb}
\usepackage[pdftex]{graphicx}
\usepackage{bm}
\usepackage[pdftex]{color}
\usepackage{cases}
\usepackage{multirow}
\usepackage{braket}
\usepackage{mathtools}
\usepackage{comment}

\newcommand{\bket}[1]{\mathinner{|{#1}\rangle\!\rangle}}
\newcommand{\bbra}[1]{\mathinner{\langle\!\langle{#1}|}}
\newcommand{\bbraket}[1]{\mathinner{\langle\!\langle{#1}\rangle}}

\def\stacksymbols #1#2#3#4{\def\theguybelow{#2}
    \def\verticalposition{\lower#3pt}
    \def\spacingwithinsymbol{\baselineskip0pt\lineskip#4pt}
    \mathrel{\mathpalette\intermediary#1}}
\def\intermediary#1#2{\verticalposition\vbox{\spacingwithinsymbol
      \everycr={}\tabskip0pt
      \halign{$\mathsurround0pt#1\hfil##\hfil$\crcr#2\crcr
               \theguybelow\crcr}}}

\usepackage[pdftex]{hyperref}

\begin{document}
%
\title{Topological phases protected by shifted sublattice symmetry in dissipative quantum systems}

\author{Makio Kawasaki}
\affiliation{Department of Applied Physics, Hokkaido University, Sapporo 060-8628, Japan}
\author{Ken Mochizuki}
\affiliation{Advanced Institute for Materials Research (WPI-AIMR), Tohoku University, Sendai 980-8577, Japan}
\affiliation{Nonequilibrium Quantum Statistical Mechanics RIKEN Hakubi Research Team, RIKEN Cluster for Pioneering Research (CPR), 2-1 Hirosawa, Wako 351-0198, Japan}
\author{Hideaki Obuse}
\affiliation{Department of Applied Physics, Hokkaido University, Sapporo 060-8628, Japan}
\affiliation{Institute of Industrial Science, The University of Tokyo, 5-1-5 Kashiwanoha, Kashiwa,Chiba 277-8574, Japan}


 \begin{abstract}
  Dissipative dynamics of quantum systems can be classified topologically based on the correspondence between the Lindbladian in the Gorini-Kossakowski-Sudarshan-Lindblad equation and  the non-Hermitian Hamiltonian in the Schr\"{o}dinger equation.
  While general non-Hermitian Hamiltonians are classified into 38 symmetry classes, previous studies have shown that the Lindbladians are classified into 10 symmetry classes due to a physical constraint.
  In this work, however, we unveil a topological classification of Lindbladians based on shifted sublattice symmetry (SLS), which can increase the number of symmetry classes for the Lindbladians.
  We introduce shifted SLS so that the Lindbladian can retain this symmetry and take on the same role as SLS for the topological classification.
  For verification, we construct a model of a dissipative quantum system retaining shifted SLS and confirm the presence of edge states protected by shifted SLS.
  Moreover, the relationship between the presence of shifted SLS protected edge states and the dynamics of an observable quantity is also discussed.

 \end{abstract}


\maketitle
\section{introduction}
\label{sec:introduction}

Recently, topological phases of non-Hermitian systems have attracted much attention \cite{esaki2011edge,lee2016anomalous,lieu2018topological,kawabata2018parity,yokomizo2019non,liu2019second,zeng2020topological,ashida2020non,bergholtz2021exceptional}.
The topological phases are classified by the presence or absence of symmetries, and non-Hermiticity of Hamiltonians enriches their symmetry class \cite{gong2018topological,kawabata2019topological,kawabata2019symmetry}.
Moreover, non-Hermitian systems exhibit unique topological phenomena that have no counterpart in closed systems, non-Hermitian skin effects, and the breakdown of the bulk-edge correspondence \cite{yao2018edge,okuma2020topological,kawasaki2020bulk,zhang2020correspondence,sone2020exceptional,borgnia2020non}.
Non-Hermitian Hamiltonians describe certain types of open systems, for example, classical optical systems with gain and/or loss \cite{weimann2017topologically,bandres2018topological} and postselected open quantum systems \cite{mochizuki2016,xiao2017observation,xiao2020non}.

Whereas the Schr\"odinger equation with a non-Hermitian Hamiltonian describes the short-time (i.e., without quantum jumps) dynamics of a dissipative quantum system, it is necessary to consider the time evolution of a density operator \(\rho\) for a long-time dynamics with quantum jumps.
For a memoryless dissipative quantum system, the Gorini-Kossakowski-Sudarshan-Lindblad (GKSL) equation \cite{breuer2002theory} \(\displaystyle i\frac{d\rho}{dt}=\hat{\mathcal{L}}[\rho]\) describes the time evolution of the density operator.
The non-Hermitian superoperator called the Lindbladian \(\hat{\mathcal{L}}\) completely characterizes the Markovian dynamics of the system.
Accordingly, topological phases associated with the GKSL equation need to be considered in order to understand the topological phenomena of dissipative quantum systems.
Nonequilibrium phenomena in dissipative many-body quantum systems have attracted attention recently \cite{landi2021non,prosen2011open,minganti2018spectral,mori2020resolving,sa2020complex,haga2021liouvillian,nakagawa2021exact}.
In particular, the topological properties of steady states \cite{diehl2011topology,bardyn2013topology,rivas2013density,budich2015dissipative,zhang2018topological,altland2021symmetry} and the Lindbladian spectra \cite{dangel2018topological,van2019dynamical,song2019non,goldstein2019dissipation,lieu2020tenfold,yoshida2020fate,huang2020quantum,longhi2020unraveling,pan2021point,flynn2021topology} have been studied.
In addition, symmetries for Lindbladians, which are important for the classification of topological phases, have been examined \cite{prosen2012p,buvca2012note,albert2014symmetries,van2018symmetry,10.21468/SciPostPhys.9.4.052,lieu2020symmetry}.

In particular, Ref.\ \cite{lieu2020tenfold} studied the topological classification of the spectra of Lindbladians by using the topological classification of non-Hermitian Hamiltonians.
The authors suggested that the Lindbladian spectra are classified into 10 symmetry classes, while non-Hermitian Hamiltonians are classified into 38 symmetry classes \cite{kawabata2019symmetry}.
The absence of some symmetry classes originates from the fact that Lindbladians cannot possess certain types of symmetries which are used in the classification of non-Hermitian Hamiltonians because the total probability of quantum states does not amplify in time. 
We remark that although more than half of the 38 symmetry classes for non-Hermitian Hamiltonians are classified by using sublattice symmetry (SLS), this symmetry was not considered in the previous work \cite{lieu2020tenfold}.

In this paper, we clarify the topological phases of Lindbladians originating from a modified version of SLS, which was not predicted in the previous tenfold classification \cite{lieu2020tenfold}.
To this end, we introduce shifted SLS which is defined by the combination of a constant shift of a Lindbladian spectrum and ordinary SLS.
Whereas Lindbladians cannot retain ordinary SLS, Lindbladians can retain shifted SLS since the constant shift prevents violating the physical constraint.
To exemplify the topological phases originating from shifted SLS, we consider a Su-Schrieffer–Heeger (SSH) model with dissipation which retains shifted SLS and confirm the emergence of topological edge states protected by shifted SLS.

This paper is organized as follows.
In Sec.\ \ref{sec:sym_lind} we review the Lindbladian spectra and their topological classification.
Then we define shifted SLS and explain that shifted SLS is a well-defined symmetry in dissipative quantum systems.
We construct an SSH model with intra-sublattice dissipation which retains shifted SLS in Sec.\ \ref{sec:ssh model}.
We show that this model has edge states protected by shifted SLS, and they are robust against symmetry-preserving disorders.
We also show that the emergence of the edge states affects the decay rate of the local occupation numbers.
We summarize the results in Sec.\ \ref{sec:summary}.

\section{Symmetry classes and topological classification of Lindbladians}
\label{sec:sym_lind}
\subsection{GKSL equation and Lindbladians}
\label{subsec:lind_eq}
We consider a quantum system coupled to environments.
Assuming that the dynamics of the system is Markovian, the reduced density operator of the system \(\rho\) obeys the GKSL equation \cite{breuer2002theory},
\begin{equation}
  i\frac{d\rho}{dt}=\hat{\mathcal{L}}[\rho]=[\mathcal{H},\rho]+i\sum_{\mu}(2L_{\mu}\rho L_{\mu}^{\dagger}-\{L_{\mu}^{\dagger}L_{\mu},\rho\}). \label{eq:lindblad}
\end{equation}
The first term describes the unitary dynamics generated by a Hermitian Hamiltonian \(\mathcal{H}\), and the second term describes the dissipative dynamics caused by non-Hermitian jump operators \(L_{\mu}\).
The superoperator \(\hat{\mathcal{L}}\) is called the Lindbladian.
Throughout this paper, symbols with a hat, such as \(\hat{O}\), represent superoperators, and symbols without a hat represent ordinary operators, matrices, or numbers.
The solution of the GKSL equation \eqref{eq:lindblad} is formally described as
\begin{equation}
  \rho(t)=\exp(-i\hat{\mathcal{L}}t)[\rho(0)], \label{eq:timeevo}
\end{equation}
in a manner similar to the Schr\"{o}dinger equation.

Since Lindbladians are non-Hermitian linear superoperators, we can consider an eigenequation,
\begin{equation}
  \hat{\mathcal{L}}[\rho_j]=\nu_j\rho_j,~\nu_j\in\mathbb{C}. \label{eq:eigen_lindblad}
\end{equation}
\(-\mathrm{Im}\,\nu_j\) corresponds to the decay rate of the eigenoperator \(\rho_j\).
To ensure that the solution of the GKSL equation given in Eq.\ \eqref{eq:timeevo} does not amplify in time, the imaginary part of the eigenvalues of general Lindbladians must not be positive \cite{pc}.

\subsection{Review of the tenfold classification of Lindbladians}
\label{subsec:class}
We briefly review the topological classification of the Lindbladian spectra studied in Ref.\ \cite{lieu2020tenfold}.
If the Hamiltonian describes noninteracting fermions and the jump operators are linear in fermionic operators, the corresponding Lindbladian also describes a noninteracting system.
Then, we can classify the Lindbladian by using the classification of noninteracting non-Hermitian Hamiltonians \cite{kawabata2019symmetry}.

First, we review the noninteracting Lindbladians and their description  \cite{prosen2008third,prosen2010spectral}.
For the dissipative noninteracting system of \(N\) fermions, the Hamiltonian and jump operators are expressed as
\begin{equation}
  \mathcal{H}=\sum_{i,j}^{2N}w_iH_{i,j}w_j,~L_{\mu}=\sum_j^{2N}l_{\mu,j}w_j,
\label{eq:H_Majorana}
\end{equation}
where Hermitian operators \(w_j\) satisfy \(\{w_i,w_j\}=2\delta_{i,j}\) and are called Majorana operators.
We note that the fermionic operators \(c\) and the Majorana operators are related as $c_j=(w_{2j-1} + i w_{2j})/2$.
In such cases, the Lindbladian can be written as
\begin{equation}
  \hat{\mathcal{L}}=2\begin{pmatrix} \hat{\bm{\psi}}^{\dagger} & \hat{\bm{\psi}} \end{pmatrix} \begin{pmatrix} -Z^T & Y \\ 0 & Z \end{pmatrix} \binom{\hat{\bm{\psi}}}{\hat{\bm{\psi}}^{\dagger}}-A_0\hat{I}. \label{eq:3quant}
\end{equation}
Physically, \(Z\) represents the damping modes, and \(Y\) determines structures of the steady state as shown in Sec.\ \ref{subsec:loc_occup}.
\(\hat{\bm{\psi}}=\begin{pmatrix} \hat{\psi}_1 & \hat{\psi}_2 & \cdots & \hat{\psi}_{2N}\end{pmatrix}\) are fermionic superoperators satisfying \(\{\hat{\psi}_i,\hat{\psi}^{\dagger}_j\}=\delta_{i,j}\), whose operation on $\rho$ are defined by
\begin{equation}
  \hat{\psi}_j[\rho]\coloneqq\frac{1}{2}(w_j\rho+\hat{P}[\rho]w_j),~\hat{\psi}_j^{\dagger}[\rho]\coloneqq\frac{1}{2}(w_j\rho-\hat{P}[\rho]w_j), \label{eq:sp_fermion}
\end{equation}
where \(\hat{P}=e^{i\pi\sum\hat{\psi}_j^{\dagger}\hat{\psi}_j}\) is the parity superoperator and \(\hat{I}\) denotes the identity superoperator.
The matrices \(Z\) and \(Y\) are defined by
\begin{equation}
    Z\coloneqq H+i\mathrm{Re}M,~Y\coloneqq2\mathrm{Im}M, \label{eq:matrices_lind}
\end{equation}
where $H$ and \(M\) are Hermitian \(2N\times2N\) matrices whose matrix elements are defined by $H_{i,j}$ in Eq.~(\ref{eq:H_Majorana}) and
\begin{equation}
  M_{i,j}\coloneqq\sum_{\mu}l_{\mu,i}l_{\mu,j}^*, \label{eq:matrix_M}
\end{equation}
respectively.
\(H\) has particle-hole symmetry \(H=-H^*\) because of the anticommutation relation of the Majorana operators.
The positive matrix $M$ describes the effects of the dissipation and is called the bath matrix.
The matrix \(Z\) becomes non-Hermitian when \(\mathrm{Re}M\ne0\).
\(A_0\) is given as \(A_0=2\mathrm{tr}[M]\).
If \(Z\) is diagonalizable, \(\hat{\mathcal{L}}\) is also diagonalizable, and the Lindbladian spectrum is completely determined by the spectrum of \(Z\):
\begin{equation}
  \hat{\mathcal{L}}=-4\sum_j^{2N}\lambda_j\hat{b}'_j\hat{b}_j, \label{eq:diagonal}
\end{equation}
where \(\lambda_j\) are eigenvalues of \(Z\).
\(\hat{b}'_j\) and \(\hat{b}_j\) are generalized fermionic superoperators satisfying \(\{\hat{b}_j,\hat{b}_k\}=\{\hat{b}'_j,\hat{b}'_k\}=0\) and \(\{\hat{b}_j,\hat{b}'_k\}=\delta_{j,k}\) \cite{prosen2008third}.

Hence, we can focus on the topological classification of the non-Hermitian matrix \(Z\) instead of $\hat{\mathcal{L}}$ in order to classify the Lindbladian spectrum.
We identify the symmetry classes of the Lindbladian through \(Z\) hereafter.
Topological classification of \(Z\) is provided in Ref.\ \cite{lieu2020tenfold}, and the definition of symmetries in the AZ and AZ$^\dagger$ classes follows from Ref.\ \cite{kawabata2019symmetry}.
Since the Lindbladian spectrum and \(Z\) are connected by Eq.\ \eqref{eq:diagonal}, \(Z\) must have eigenvalues with positive imaginary parts.
Due to this constraint, \(Z\) cannot have symmetries in the AZ class and SLS.
For example, if \(Z\) has time-reversal symmetry in the AZ class, a unitary operator \(\mathcal{T}\) exists such that \(Z=\mathcal{T}^{-1}Z^*\mathcal{T}\).
This equation ensures that \(Z\) has complex-conjugate eigenvalue pairs.
Then \(Z\) has eigenvalues with negative imaginary parts if \(Z\) has eigenvalues with finite imaginary parts, which conflicts with the discussion in the previous subsection.
Similarly, we can show that \(Z\) cannot retain SLS.
If \(Z\) has SLS, a unitary operator \(\mathcal{S}\) exists such that
\begin{equation}
Z=-\mathcal{S}^{-1} Z \mathcal{S},~ \mathcal{S}^2=1.
\label{eq:SLS}
\end{equation}
This equation ensures that \(Z\) has the eigenvalues pairs \(\pm\lambda\) and is a traceless matrix.
This also conflicts with the physical constraint of the Lindbladian spectrum.
We can show that other symmetries in the AZ class also generate eigenvalues with negative imaginary parts; thus, AZ symmetry is prohibited for \(Z\).
Consequently, \(Z\) can have symmetries only in the AZ\(^{\dagger}\) class, and \(Z\) is classified in tenfold symmetry classes, which is the same number as in the case of Hermitian Hamiltonians.

\subsection{Shifted sublattice symmetry}
\label{subsec:passive_sls}
While Lindbladians cannot retain ordinary SLS as explained below Eq.\ \eqref{eq:SLS}, we will show that a modified version of SLS, called shifted SLS, is permissible for the Lindbladians.
We will also show that shifted SLS is topologically equivalent to ordinary SLS.

Here, we introduce shifted SLS for a Lindbladian, which is the combination of a constant shift of the decay rate and ordinary SLS.
Shifted SLS is defined for the traceless part of \(Z\),
\begin{equation}
  Z'\coloneqq Z-iaI,~a\coloneqq \frac{\mathrm{tr}[Z]}{i\mathrm{tr}[I]}, \label{eq:passive_SLSb}
\end{equation}
and the definition of shifted SLS is given by
\begin{align}
  Z'=-\mathcal{S}Z'\mathcal{S}^{-1},~\mathcal{S}^2=I, \label{eq:passive_SLS}
\end{align}
in real space and
\begin{equation}
  Z'(\bm{k})=-\mathcal{S}Z'(\bm{k})\mathcal{S}^{-1},~Z'(\bm{k})\coloneqq Z(\bm{k})-iaI, \label{eq:passive_SLSm}
\end{equation}
in momentum space, where \(I\) denotes the identity matrix whose dimension is the same as \(Z\).
Since \(H\) is traceless and \(M\) is positive by definition, \(a\) always satisfies
\begin{equation}
  a=\frac{i\mathrm{tr}[\mathrm{Re}M]}{i\mathrm{tr}[I]}=\frac{\mathrm{tr}[M]}{\mathrm{tr}[I]}\geq 0 \label{eq:a_sls}.
\end{equation}
Shifted SLS ensures that \(Z\) has eigenvalue pairs \(\pm\lambda+ia\).
By construction, the imaginary part of all eigenvalues of shifted SLS symmetric \(Z\) is positive, which satisfies the physical constraint for Lindbladians.
Hence, shifted SLS can be retained for Lindbladians.

We remark that shifted SLS imposes the same topological constraints as ordinary SLS because the shifted SLS symmetric matrix can be continuously deformed into an SLS symmetric matrix without closing the gap and changing the symmetry class.
Furthermore, \(Z'\) must inherit the same symmetries if \(Z\) retains some symmetries of the AZ\(^{\dagger}\) class and vice versa.
In other words, the topological phase of \(Z'\) corresponds to that of \(Z\).
Shifted SLS enables us to explore symmetry classes of Lindbladians which were not studied in Ref.\ \cite{lieu2020tenfold}, that is, the AZ\(^\dagger\) class with shifted SLS.
Remarkably, among 38 fold classifications, 22 symmetry classes are related to SLS.
This is the main result of the present work.

We also note that topological phases associated with pseudo-Hermiticity in Lindbladians can be realized in a way similar to shifted SLS.
While ordinary pseudo-Hermiticity \(Z=\eta Z^{\dagger} \eta^{-1}\) conflicts with the physical constraint for Lindbladians, shifted pseudo-Hermiticity
\begin{equation}
  Z'=\eta Z'^{\dagger} \eta^{-1},~\eta^2=I \label{eq:passive_pH}
\end{equation}
is a valid symmetry for Lindbladians.
We can investigate the topological phases protected by shifted pseudo-Hermiticity in dissipative quantum systems.
In addition, \(\mathcal{PT}\) symmetry, closely related to pseudo-Hermiticity, for Lindbladians was already defined in a shifted way in Ref.\ \cite{prosen2012p}.

\section{Example : SSH model with intrasublattice dissipation}
\label{sec:ssh model}
In this section, we consider an SSH model with dissipation which retains shifted SLS and study its topological phase.
In Sec.\ \ref{subsec:model} we explain the details of our model.
We calculate topological invariants based on shifted SLS in Sec.\ \ref{subsec:topoinv}.
We numerically confirm the emergence of edge states and study their robustness against symmetry-preserving perturbation in Sec.\ \ref{subsec:bec}.
Finally, we investigate how edge states protected by shifted SLS affect the dynamics of the local occupation number in Sec.\ \ref{subsec:loc_occup}.

\subsection{Model and symmetry}
\label{subsec:model}
\begin{figure}[tb]
  \centering
  \includegraphics[width=\columnwidth]{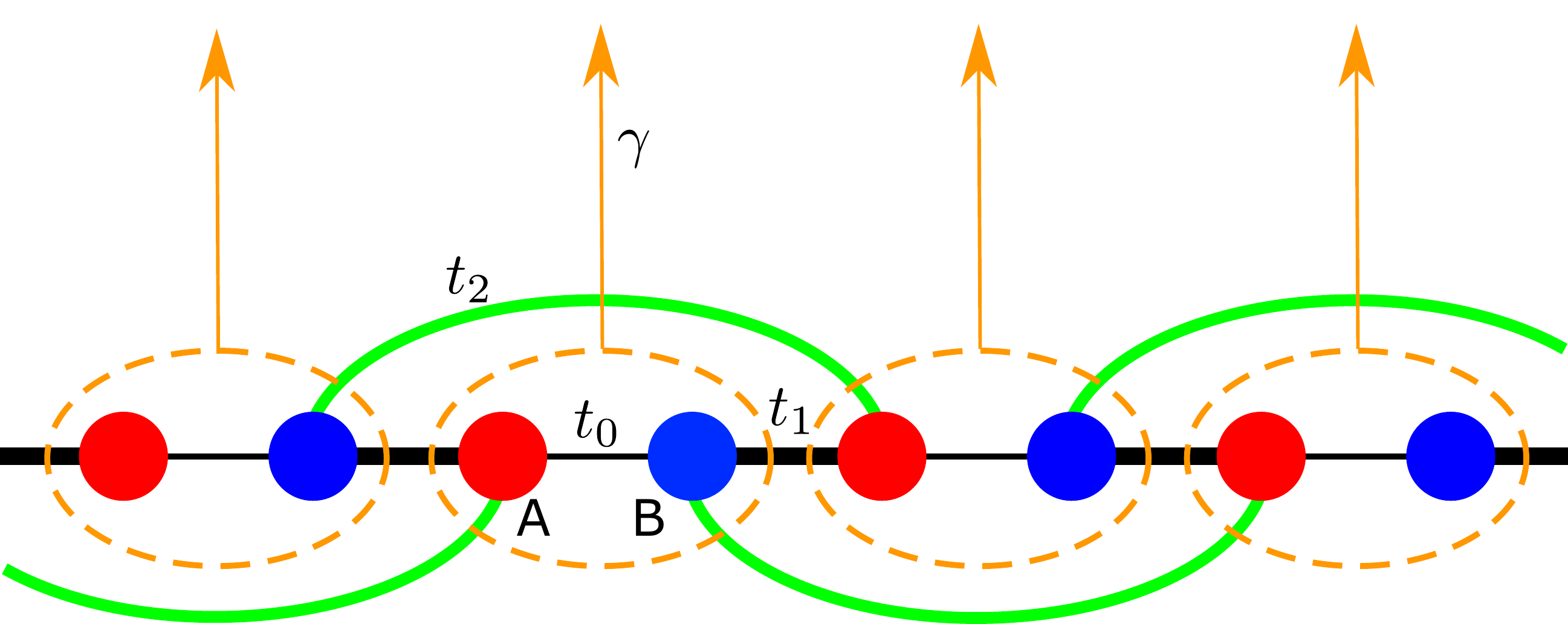}
  \caption{The schematic of the SSH model with intrasublattice dissipation.
  The black thin lines represent the intrasublattice hoppings.
  The black and green thick lines represent the hopping to the first- and second-nearest-neighbor unit cells, respectively.
  The orange arrows correspond to the dissipation process defined in Eq.\ \eqref{eq:jump}.}
  \label{fig:ssh}
\end{figure}

We consider the SSH model \cite{su1980soliton} with second nearest-neighbor coupling and dissipation, whose Lindbladian retains shifted SLS.
The Hamiltonian is given by
\begin{equation}
  \mathcal{H}=\sum_x\sum_{j=0}^{2}(t_jc_{x,B}^{\dagger}c_{x+j,A}+\mathrm{H.c.}),~t_j\in\mathbb{R}, \label{eq:ssh}
\end{equation}
where \(c_{x,s}\) denotes the fermionic annihilation operator at position \(x\) with sublattice index \(s\in\{A,B\}\).
The jump operator acting on the unit cell is also given as one-body losses of intra-sublattice fermions:
\begin{equation}
  L_x=\gamma(c_{x,A}+c_{x,B}). \label{eq:jump}
\end{equation}
Therefore, the jump operator describes the dissipation of an ``in-phase'' pair of the fermions in each unit cell.
The non-negative parameter \(\gamma\) denotes the strength of the dissipation.
Figure \ref{fig:ssh} represents the schematic of our model.
The coupling strength between the system and environments does not depend on the position and the sublattice index.
According to Sec.\ \ref{subsec:class}, we obtain matrices \(H,~M\), and \(Z\) by rewriting the fermionic annihilation operators at site $x$ on sublattice $s$ as the combination of Majorana operators \(c_{x,s}=(w_{x,s,\alpha}+iw_{x,s,\beta})/2\).
The Majorana representation of the SSH model Hamiltonian \(H\) is interpreted as two independent Kitaev chains as
\begin{equation}
  H=\begin{pmatrix}
  H^{\mathrm{kit}} & 0 \\ 0 & -H^{\mathrm{kit}}
\end{pmatrix}, \label{eq:hamiltonian_majorana}
\end{equation}
where \(H^{\mathrm{kit}}\) is the Majorana representation of the Kitaev chain Hamiltonian \cite{kitaev2001unpaired},
\begin{equation}
  H^{\mathrm{kit}}=\sum_x\sum_{j=0}^2 \left(\frac{i}{4}t_j\ket{x,u}\bra{x+j,\bar{u}}+\mathrm{H.c.}\right). \label{eq:kitaev}
\end{equation}
The upper and lower blocks in Eq.\ \eqref{eq:hamiltonian_majorana} are spanned by the bases \(\{\ket{xA\beta},\ket{xB\alpha}\}\) and \(\{\ket{xA\alpha},\ket{xB\beta}\}\), respectively.
\(u\) in Eq.\ \eqref{eq:kitaev} denotes the internal and sublattice indices of the Majorana operators; \(u=A\beta\) in the upper block, and \(u=A\alpha\) in the lower block.
\(\bar{u}\) means that two indices in \(u\) are flipped.
In the same basis of Eq.\ \eqref{eq:hamiltonian_majorana}, the matrix \(M\) defined in Eq.\ \eqref{eq:matrix_M} is derived as
\begin{equation}
  M=\frac{\gamma^2}{4}\sum_x\ket{x}\bra{x}\otimes\begin{pmatrix}
  I-\sigma_y & \sigma_x+i\sigma_z \\
  \sigma_x-i\sigma_z & I+\sigma_y
\end{pmatrix},
\end{equation}
where \(\sigma_{x,y,z}\) are Pauli matrices acting on the sublattice and Majorana indices.
Then we obtain the matrix \(Z\) as
\begin{equation}
  Z=\begin{pmatrix} H^{\mathrm{kit}} & 0 \\ 0 & -H^{\mathrm{kit}} \end{pmatrix}+\frac{i\gamma^2}{4}\sum_x\ket{x}\bra{x}\otimes \begin{pmatrix} I_2 & \sigma_x \\ \sigma_x & I_2 \end{pmatrix}. \label{eq:Z}
\end{equation}
Since dissipation realizes the anti-Hermitian coupling between two Kitaev chains in addition to the onsite dissipation, the model can be interpreted as a ladder of Kitaev chains with non-Hermitian couplings.

\begin{figure}[tb]
  \centering
  \includegraphics[width=\columnwidth]{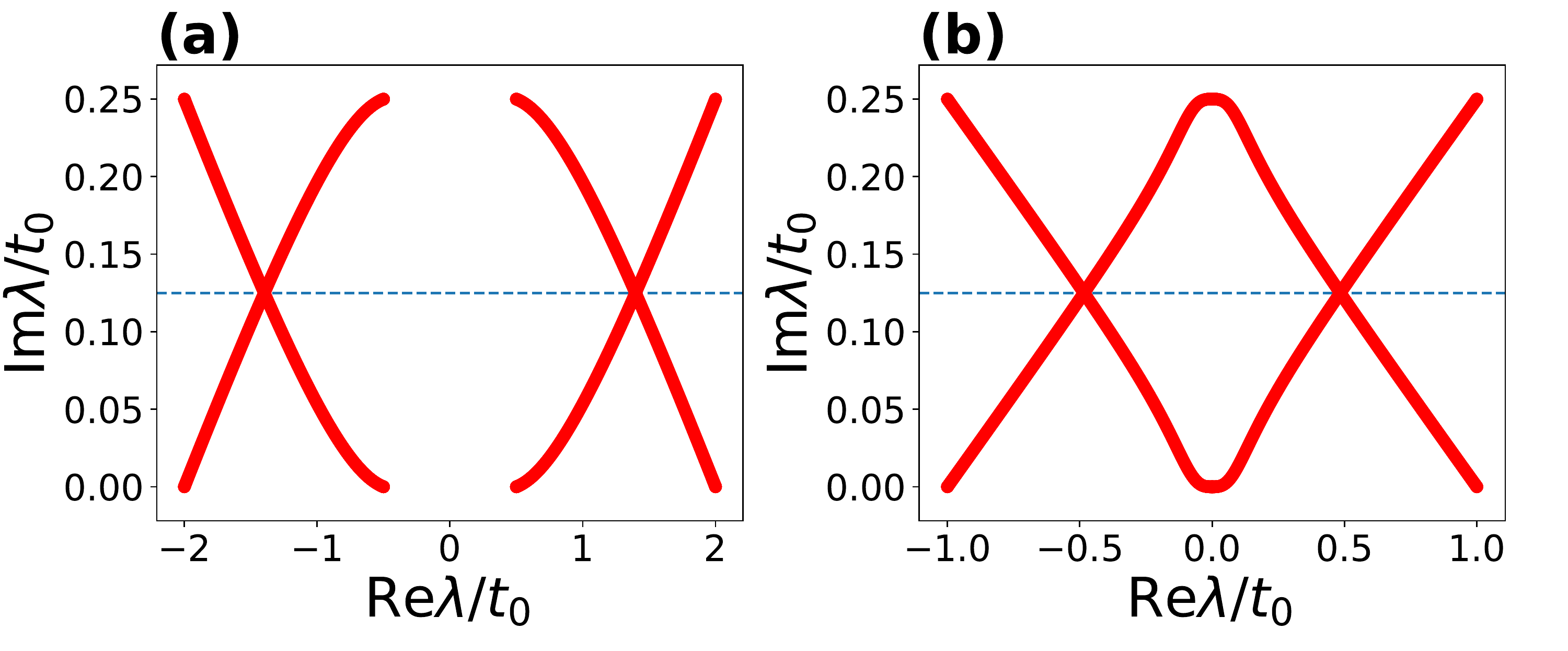}
  \caption{Spectra of \(Z\) with periodic boundary conditions with (a) \(t_1=5t_0,~t_2=2t_0\), and \(\gamma^2=0.5t_0\) and
  (b) \(t_1=2t_0,~t_2=t_0\), and \(\gamma^2=0.5t_0\).
  The dashed lines in (a) and (b) indicate \(\displaystyle \frac{ia}{t_0}=\frac{\gamma^2}{4t_0}\).}
  \label{fig:spectra_periodic}
\end{figure}

Under the periodic boundary conditions, the Fourier transform over the position space leads to $Z(k)$ in the momentum space as
\begin{equation}
  Z(k)=\frac{i}{4}\begin{pmatrix} \gamma^2 & -t_k^* & 0 & \gamma^2 \\
  t_k & \gamma^2 & \gamma^2 & 0 \\
  0 & \gamma^2 & \gamma^2 & t_k^* \\
  \gamma^2 & 0 & -t_k & \gamma^2
\end{pmatrix},t_k=\sum_{j=0}^2 e^{ikj}t_j. \label{eq:Z_k}
\end{equation}
The eigenvalues of \(Z(k)\) are calculated as
\begin{equation}
  \lambda_{p,q}(k)=\frac{i\gamma^2}{4}+\frac{(-1)^q}{4}\sqrt{|t_k|^2-\gamma^4+(-1)^{p+1}2i\gamma^2\mathrm{Re}t_k},\label{eq:eigval}
\end{equation}
where \(p,q\in\{0,1\}\).
\(Z(k)\) has a real line gap, as shown in Fig. \ref{fig:spectra_periodic}(a), in a certain parameter region where topological phases are well defined.
The gap closes for \(\mathrm{Re}t_k=0\) and \(|t_k|^2\leq\gamma^4\).
These conditions are satisfied in Fig.\ \ref{fig:spectra_periodic}(b), and the real line gap of \(Z\) closes.

According to Sec.\ \ref{subsec:passive_sls}, we focus on the traceless part of \(Z(k)\),
\begin{equation}
  Z'(k)\coloneqq Z(k)-\frac{i\gamma^2}{4}I_4. \label{eq:Z_prime}
\end{equation}
\(Z'(k)\) [or, equivalently, \(Z(k)\)] has all symmetries of the AZ\(^{\dagger}\) class, i.e., time-reversal, particle-hole, and chiral symmetries,
\begin{align}
  Z'(k)&=\mathcal{T}Z'^T(-k)\mathcal{T}^{-1},~\mathcal{T}=\sigma_x\otimes I_2, \label{eq:Z_trs}\\
  Z'(k)&=-Z'^*(-k), \label{eq:Z_phs}\\
  Z'(k)&=-\Gamma Z'^{\dagger}(k)\Gamma^{-1},~\Gamma=\mathcal{T}, \label{eq:Z_cs}
\end{align}
respectively.
Moreover, \(Z(k)\) has shifted SLS,
\begin{equation}
  Z'(k)=-\mathcal{S}Z'(k)\mathcal{S}^{-1},~\mathcal{S}=I_2\otimes\tau_z. \label{eq:Z_sls}
\end{equation}
The Pauli matrices \(\tau_{x,y,z}\) act on the block matrix spaces in Eq.\ \eqref{eq:Z}.
The symmetry operator of shifted SLS commutes with those of time-reversal symmetry in Eq.\ \eqref{eq:Z_trs} and particle-hole symmetry in Eq.\ \eqref{eq:Z_phs}.
According to the topological classification for non-Hermitian Hamiltonians \cite{kawabata2019symmetry}, this model belongs to class BDI\(^{\dagger}+\mathcal{S}_{++}\) (which is topologically equivalent to class BDI\(~+\mathcal{S}_{++}\) ), whose topological phase is classified in \(\mathbb{Z}\oplus\mathbb{Z}\).
This topological phase is not included in the prior classification of Lindbladians \cite{lieu2020tenfold} because their classification misses the contribution of shifted SLS.

\subsection{Topological invariant}
\label{subsec:topoinv}
\(\mathbb{Z}\oplus\mathbb{Z}\) topological invariants can be calculated as follows \cite{kawabata2019symmetry}.
\(Z'(k)\) is pseudo-Hermitian by combining SLS and chiral symmetry,
\begin{equation}
   Z'(k)=\eta Z'^{\dagger}(k)\eta^{-1},~\eta=\Gamma\mathcal{S}=\sigma_x\otimes\tau_z. \label{eq:Z_pseudoH}
\end{equation}
Then \(Z'(k)\) has the block-diagonal form \(Z'(k)=\mathrm{diag}\{Z'_+(k),Z'_-(k)\}\), and each subspace is spanned by the biorthogonal vectors \(\ket{\phi_n^{\pm}}\) and \(\bket{\phi_n^{\pm}}\).
These vectors are defined by
\begin{equation}
  \eta\ket{\phi_n^{\pm}}=\pm\bket{\phi_n^{\pm}}. \label{eq:subsector}
\end{equation}
Since \(Z'_{\pm}(k)\) has SLS \(Z'_{\pm}(k)=-\mathcal{S}_{\pm}Z'_{\pm}(k)\mathcal{S}_{\pm}^{-1}\), we can calculate the winding numbers
\begin{equation}
  w_{\pm}\coloneqq\frac{1}{4\pi i}\int_{\mathrm{BZ}}dk\mathrm{tr}\left[\mathcal{S}_{\pm}Z'^{-1}_{\pm}(k)\frac{dZ'_{\pm}(k)}{dk}\right]  \label{eq:topoinv_winding}
\end{equation}
for each subsector.

\begin{figure}[tb]
  \centering
  \includegraphics[width=\columnwidth]{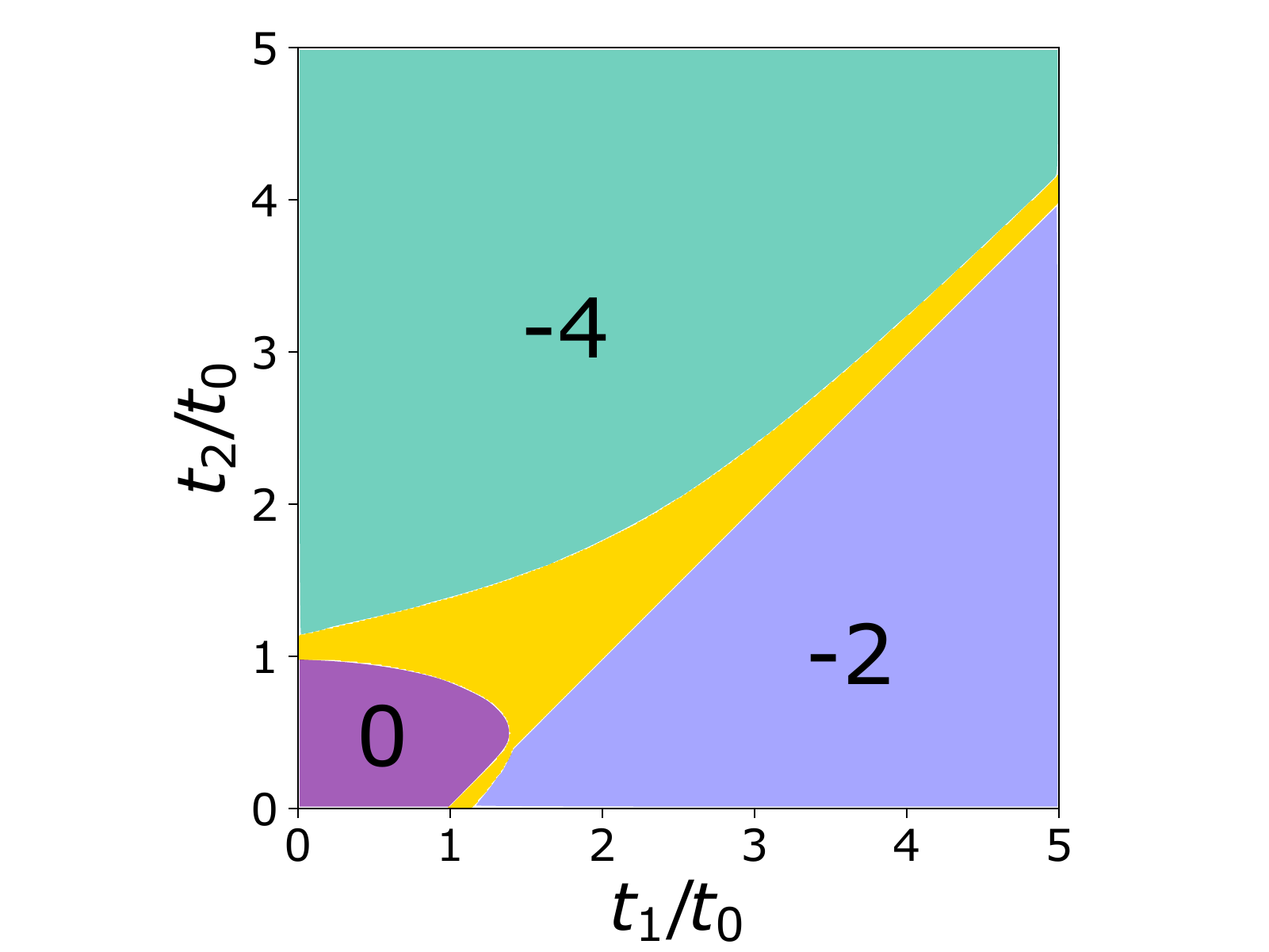}
  \caption{Phase diagram of the topological invariant $w$ for the dissipative SSH model with the second nearest neighbor coupling.
  We set the strength of dissipation to \(\gamma^2=0.5t_0\).
  In the yellow region, the real line gap closes, and the winding number cannot be defined.}
  \label{fig:phasediag}
\end{figure}

We find that two topological invariants coincide in our model; a detailed discussion is provided in Appendix \ref{app:topoinv}.
The topological invariant is given as
\begin{equation}
  w=\frac{1}{2\pi i}\int_{\mathrm{BZ}}dk\left[\frac{d}{dk}\ln(t^*_k-i\gamma^2)+\frac{d}{dk}\ln(t^*_k+i\gamma^2)\right]. \label{eq:winding_ssh}
\end{equation}
\(w\) corresponds to the sum of the winding number of two complex functions, \(t^*_k\pm i\gamma^2\).
The dependence of the topological invariant \(w\) on the parameters \(t_1\) and \(t_2\) is shown in Fig.\ \ref{fig:phasediag}.

\subsection{Bulk-edge correspondence}
\label{subsec:bec}
\begin{figure}[tbp]
  \centering
  \includegraphics[width=\columnwidth]{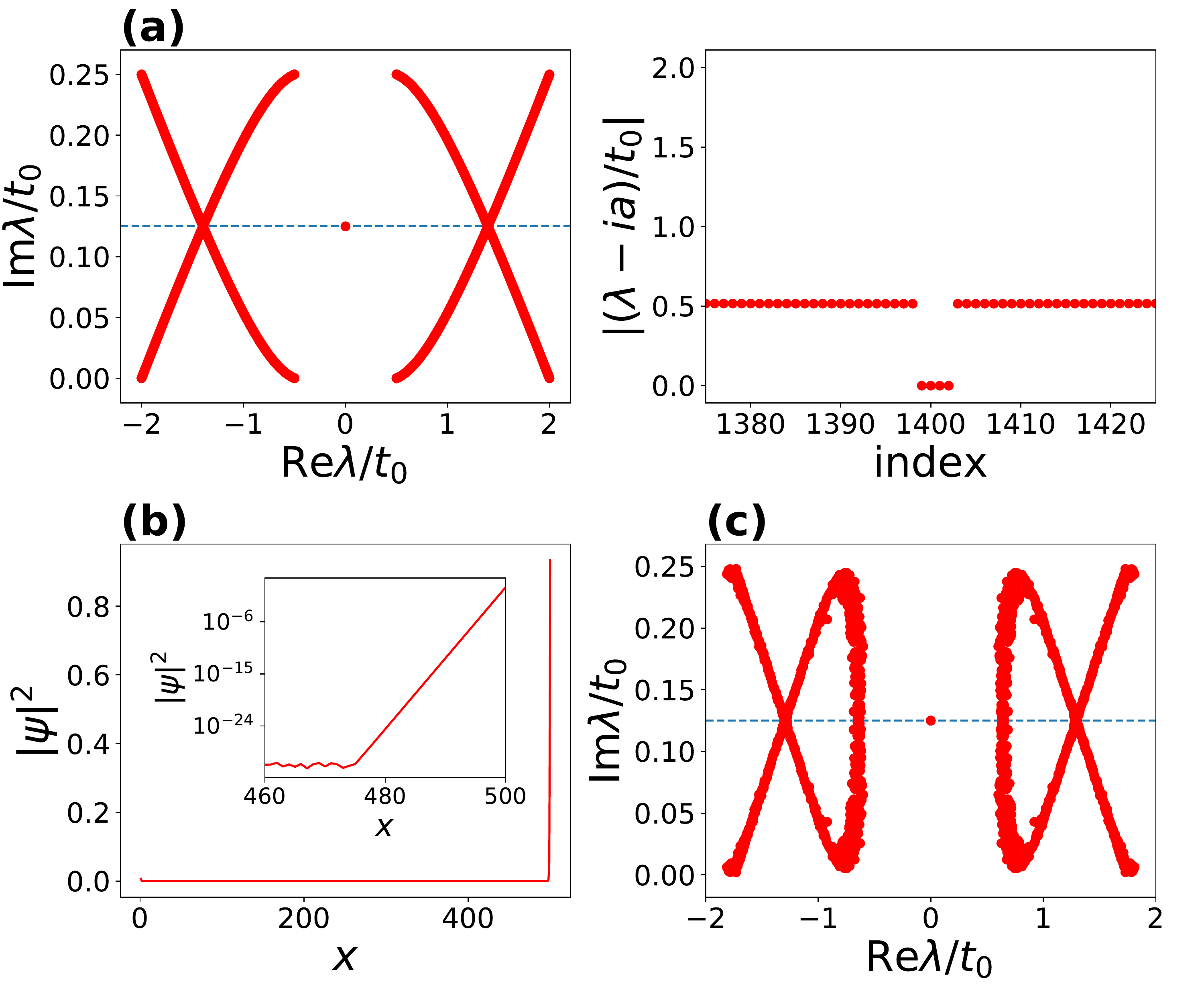}
  \caption{(a) The spectrum of \(Z\) with open boundary conditions with \(N=500,~t_1=5t_0,~t_2=2t_0\), and \(\gamma^2=0.5t_0\).
  Left: The spectrum on a complex plane.
  Right: The absolute value of the spectrum near \(ia=i\gamma^2/4\). The horizontal axis represents the order of the eigenvalues.
  The four degenerated eigenstates appear at $\lambda=i\gamma^2/4.$
  (b) The spatial distribution of an edge state.
  We plot \(|\psi|^2\) as \(|\psi_x|^2\coloneqq \sum_{s=A,B}\sum_{\mu=\alpha,\beta}|\psi_{x,s,\mu}|^2\).
  The inset shows the semi-logarithmic plot near \(x=500\).
  (c) The spectrum of \(Z\) with random hopping with
  \(\bar{t}_1=2t_0,~\bar{t}_2=4t_0\), and \(\gamma^2=0.5t_0\).
  The dashed lines in (a), and (c) indicate \(\displaystyle \frac{ia}{t_0}=\frac{\gamma^2}{4t_0}\).}
  \label{fig:spectrum}
\end{figure}

In this section, we numerically diagonalize \(Z\) and show the existence of the edge states \cite{state}.
We impose open boundary conditions on the Hamiltonian in Eq.\ \eqref{eq:ssh}, and correspondingly, there are two boundaries at each end in the system.
Figure \ref{fig:spectrum} shows spectra and edge states of \(Z\) with open boundary conditions.
We observe four edge states in Fig.\ \ref{fig:spectrum}(a) whose parameters give $w=-2$ when periodic boundary conditions are imposed on the system.
Because of the two boundaries of the system, this result verifies the bulk-edge correspondence.
Since \(Z\) and \(Z'\) are related as in Eq.\ \eqref{eq:Z_prime} and the eigenvalue of the edge states of \(Z'\) is zero due to SLS, the edge states are the eigenstates of \(Z\) with eigenvalue \(\frac{i\gamma^2}{4}\).
That is, the eigenvalues of edge states are topologically protected by shifted SLS as well as the existence of the edge states.
In fact, the eigenvalues of the edge states of noninteracting Lindbladians with shifted SLS are generally degenerate when the topological invariant is larger than 1.
The edge states are localized near boundaries, as shown in Fig.\ \ref{fig:spectrum}(b).
We have numerically verified that the number of edge states equals \(2|w|\) in the entire parameter region in Fig.\ \ref{fig:phasediag}.

We also study the robustness of the edge states.
We impose spatial disorder on hopping amplitudes.
The hopping amplitude between \((x,B)\) and \((x+j,A)\) is given as \(t_{x,j}=\bar{t}_j+\delta t_{x,j}~(j=0,1,2)\), where \(\bar{t}_j\) is a mean value of \(t_{x,j}\) and \(\delta t_{x,j}\) is the uncorrelated spatial randomness that obeys the uniform distribution in the range \([-0.3t_0,0.3t_0]\).
The edge states will be robust against the disorder because the disorder does not break any symmetry on \(Z\) including shifted SLS.
Figure \ref{fig:spectrum}(c) shows the spectrum of \(Z\) with the disorder.
In this parameter \(w=-4\), and eight edge states should appear.
In Fig.\ \ref{fig:spectrum}(c) eight edge states with eigenvalue \(\frac{i\gamma^2}{4}\) appear, and the edge states are robust despite the bulk states being perturbed by the disorder.
Hence, we verify that the winding number originating from shifted SLS corresponds to the number of topological edge states, and the eigenvalues of the edge states are topologically protected.
We note that the edge states may not be robust against the spatial disorder on \(\gamma\).
Although such disorder does not break any AZ\(^\dagger\) symmetry, the disorder breaks shifted SLS and changes the symmetry class of \(Z\).


\subsection{Dynamics of local occupation number}
\label{subsec:loc_occup}
Here, we clarify how the edge states in \(Z\) affect the time evolution of the density operator.
Since the physical meaning of fermionic superoperator \(\hat{\psi}\) defined in Eq.\ \eqref{eq:sp_fermion} is unclear, we focus on the dynamics of the local occupation number \cite{van2019dynamical}, defined as
\begin{equation}
  n_{x,s}(t)\coloneqq\mathrm{tr}[\rho(t)c^{\dagger}_{x,s}c_{x,s}],\label{eq:loc_occup}
\end{equation}
where \(c_{x,s}\) is the system's fermion in Eq.\ \eqref{eq:ssh}.

First, we derive an analytical solution of the local occupation number \(n_{x,s}(t)\).
The local occupation number \(n_{x,s}(t)\) can be rewritten as
\begin{equation}
  n_{x,s}(t)=\frac{1}{2}-\frac{i}{2}\mathrm{tr}[\rho(t)w_{x,s,\alpha}w_{x,s,\beta}]. \label{eq:loc_occup_Majorana}
\end{equation}
For noninteracting systems, the covariance matrix defined by \(C_{j,k}(t)\coloneqq\mathrm{tr}[\rho(t)w_jw_k]-\delta_{j,k}\) obeys
\begin{equation}
  \frac{dC}{dt}=4i(Z^TC+CZ+Y). \label{eq:dynamics_cov}
\end{equation}
If the system reaches a steady state (\(\frac{d\rho}{dt}=0\)), the covariance matrix of the steady state \(C^{\mathrm{ss}}\) satisfies \cite{prosen2010spectral}
\begin{equation}
  Z^TC^{\mathrm{ss}}+C^{\mathrm{ss}}Z=-Y. \label{eq:cov_ss}
\end{equation}
Focusing on the discrepancy with the steady state \(\tilde{C}(t)=C(t)-C^{\mathrm{ss}}\), \(\tilde{C}(t)\) is explicitly expressed as
\begin{align}
  \tilde{C}(t)=e^{4iZ^Tt}\tilde{C}(0)e^{4iZt}. \label{eq:cov_sol}
\end{align}
Inserting this solution into Eq.\ \eqref{eq:loc_occup_Majorana}, we obtain the analytical solution of \(n_{x,s}(t)\).
In particular, we focus on the discrepancy with the steady state \(n_{x,s}(t)-n_{x,s}^{\mathrm{ss}}\), where \(n_{x,s}^{\mathrm{ss}}\coloneqq \mathrm{tr}[\rho^{\mathrm{\mathrm{ss}}}c^{\dagger}_{x,s}c_{x,s}]\) and \(\mathcal{L}[\rho^{\mathrm{ss}}]=0\).
This quantity is provided as
\begin{align}
  n_{x,s}(t)-n_{x,s}^{\mathrm{ss}}&=-\frac{i}{2}\mathrm{tr}[\rho(t)w_{x,s,\alpha}w_{x,s,\beta}]+\frac{i}{2}\mathrm{tr}[\rho^{\mathrm{ss}}w_{x,s,\alpha}w_{x,s,\beta}] \nonumber\\
  &=-\frac{i}{2}[\tilde{C}(t)]_{xs\alpha,xs\beta}. \label{eq:loc_occup_anly}
\end{align}

\begin{figure}[tbp]
  \centering
  \includegraphics[width=\columnwidth]{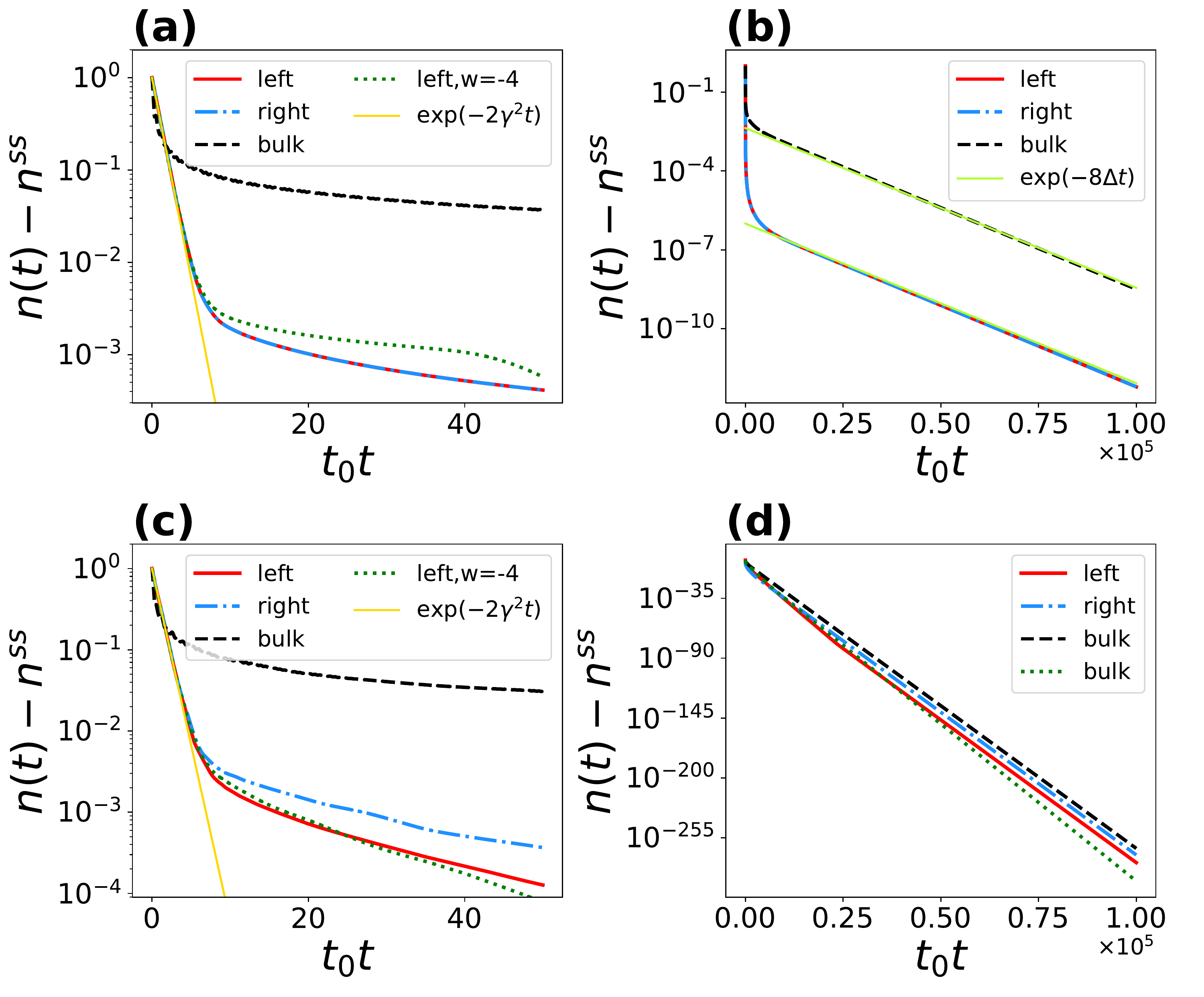}
  \caption{Time evolution of the local occupation number \(n_{x,s}(t)\).
  The labels (left, right, and bulk) in each panel denote the local occupation number at \((x,s)=(1,A),~(N,B)\), and \((\frac{N}{2},A)\), respectively.
  We set \(N=100, t_1=5t_0,~t_2=2t_0\), and \(\gamma^2=0.5t_0\)  \((w=-2)\), except for ``left, \(w=-4\)'' in (a) and (c).
  The initial states for each label are \(\rho_{in}=\ket{x,s}\bra{x,s}\), where \(\ket{x,s}\coloneqq c_{x,s}\ket{\mathrm{vac}}\).
  (a) Short-time dynamics. We set \(t_1=2t_0,~t_2=4t_0\) \((w=-4)\) for the green dotted line.
  (b) Long-time dynamics. \(\Delta=1.8\times10^{-5}t_0\) in this case.
  (c) Short-time dynamics with disorder.
  (d) Long-time dynamics with disorder.
  The green dotted line is calculated with different configurations of disorder.
  In (c) and (d), we induce the disorder in the same way as in Sec. \ref{subsec:bec}.}
  \label{fig:occup}
\end{figure}

Figure \ref{fig:occup} shows the time evolution of the local occupation number \(n_{x,s}(t)\).
We note that the local occupation number of the steady state \(n_{x,s}^{\mathrm{ss}}\) equals zero for any \(x\) and \(s=A,B\) in the model.
If an eigenstate has predominantly large overlap with \(\rho(t)\), the local occupation number should decay as
\begin{equation}
  n_{x,s}(t)-n_{x,s}^{\mathrm{ss}}\propto e^{-8\mathrm{Im}\tilde{\lambda}t}, \label{eq:n_decay}
\end{equation}
where \(\tilde{\lambda}\) denotes the corresponding eigenvalue of the eigenstate.
Since edge states, if they exist, are localized near the boundary, they have a large overlap with \(\rho_{in}\), and the short-time dynamics of \(n_{1,A}(t)\) and \(n_{N,B}(t)\) clearly reflect the lifetime of the edge states.
In those cases \(\tilde{\lambda}=\frac{i\gamma^2}{4}\), and they decay as
\begin{equation}
  n_{\mathrm{edge}}(t)-n_{\mathrm{edge}}^{\mathrm{ss}}\propto e^{-2\gamma^2t}. \label{eq:n_decay_edge}
\end{equation}
Since \(n_{\mathrm{edge}}^{\mathrm{ss}}=0\), the edge states disappear in the long-time limit, while the existence is topologically protected.
The decay rate of \(n_{\mathrm{edge}}\) depends only on the presence of edge states and not on the number of edge states or detailed values of hopping amplitudes.
As shown later, this behavior is derived from shifted SLS.
In Fig.\ \ref{fig:occup}(a), the local occupation numbers near the boundaries have the same decay rate regardless of the positions and the number of edge states, which is consistent with Eq.\ \eqref{eq:n_decay_edge}.
In contrast to short-time dynamics, the decay rate of long-time dynamics is determined from the spectral gap
\begin{equation}
  \mathrm{Im}\tilde{\lambda}=\Delta\coloneqq \min_i\mathrm{Im}\lambda_i \label{eq:spectralgap}
\end{equation}
since other states (including edge states) decay exponentially faster.
As a consequence, \(n_{x,s}(t)\) have the same decay rate, \(8\Delta=1.4\times10^{-4}t_0~(\Delta=1.8\times10^{-5}t_0)\), irrespective of their position or sublattice index, as shown in Fig.\ \ref{fig:occup}(b).

Since not only the existence but also the lifetime of the edge states is protected by shifted SLS, the decay rate of \(n_{1,A}(t)\) and \(n_{N,B}(t)\) is robust for symmetry-preserving disorders.
This result is presented in Fig.\ \ref{fig:occup}(c).
The decay rate of the edge occupation numbers obeys Eq.\ \eqref{eq:n_decay_edge} even in the presence of randomness.
On the other hand, the spectral gap defined in Eq.\ \eqref{eq:spectralgap} is not topologically protected.
We plot the long-time dynamics of \(n_{x,s}(t)\) with disorder in Fig.\ \ref{fig:occup}(d).
The spectral gaps with different disorder configurations, plotted as black and green lines in Fig.\ \ref{fig:occup}(d), are \(\Delta=5.8\times10^{-4}t_0~\mathrm{and}~7.5\times10^{-4}t_0\), respectively.
Correspondingly, bulk occupation numbers have different decay rates depending on the value of random hopping amplitudes, as shown in Fig.\ \ref{fig:occup}(d)
We conclude that the decay rate for long-time dynamics depends on the configuration of hopping amplitudes.

We have seen that the lifetime of the edge occupation number in short-time dynamics is given by the function of \(\gamma\) and does not depend on the realization of the hopping amplitude \(\{t_{x,j}\}\), while the lifetimes of bulk states depend on the hopping amplitude.
In general, the lifetime of the edge occupation number in the presence of edge states only depends only on the parameters of dissipation if the Lindbladian retains shifted SLS and particle-hole symmetry in the AZ\(^\dagger\) class.
The eigenvalues of edge states are given by \(ia = i\mathrm{tr}[M]/\mathrm{tr}[I]\) from Eq.\ \eqref{eq:a_sls}.
Since \(M\) and then \(a\) are determined only by the jump operators, the eigenvalues of edge states do not depend on the parameters of \(H\).
We also note that the decay rates of the edge occupation numbers \(n_{\mathrm{edge}}\) for both ends coincide if the system has shifted SLS (cf. Ref.\ \cite{van2019dynamical}), since shifted SLS guarantees the degeneracy of the eigenvalues of edge states and then \(\tilde{\lambda}\) in Eq.\ \eqref{eq:n_decay} coincides for both ends.

\section{Summary}
\label{sec:summary}

In this work, we have introduced shifted SLS for Lindbladians and investigated the topological phase associated with shifted SLS.
As a consequence, we have uncovered different symmetry classes of Lindbladians, that is, AZ\(^{\dagger}\) with shifted SLS classes.
Therefore, shifted SLS can increase the number of symmetry classes of Lindbladians.
We have constructed a dissipative version of the SSH model retaining shifted SLS.
We have clarified that the presence of edge states and their eigenvalues are protected by shifted SLS.
We have also studied the dynamics of the local occupation numbers and clarified that the edge occupation numbers have a universal decay rate which is also protected by shifted SLS for short-time dynamics.
Our work opens up an arena to study nonequilibrium topological phenomena in dissipative quantum systems.

\section*{Acknowledgments}

We thank Y.\ Asano, M.\ Sato, and K.\ Yakubo for helpful discussions.
M.\ K.\ was supported by JST SPRING (Grant No.\ JPMJSP2119).
This work was also supported by JST ERATO-FS Grant No. JPMJER2105 and KAKENHI (Grants
No.\ JP18J20727, No.\ JP18H01140, No.\ JP19K03646, No.\ 20H01828, No.\ 20H01828,  No.\ JP20H01845, No.\ JP21H01005, and No.\ JP22H01140).

\appendix
\section{Conditions of the jump operators for the matrix \(Z\) to retain shifted SLS}
\label{app:sls}
In this appendix, we clarify the sufficient conditions for retaining shifted SLS in one-dimensional (1D) dissipative systems.
First of all, from the Hermitian part of Eq.\ \eqref{eq:passive_SLS}, a Hermitian Hamiltonian (under the Majorana representation) \(H\) must have sublattice symmetry \(H(k)=-\mathcal{S}H(k)\mathcal{S}^{-1}\).
We assume a sublattice symmetric 1D Hamiltonian with two sublattices.
We choose the symmetry operator \(\mathcal{S}\) as \(\mathcal{S}=I_2\otimes\sigma_z\) without loss of generality by applying an appropriate unitary transformation.

Focusing on the anti-Hermitian part of Eq.\ \eqref{eq:passive_SLSm},
\begin{equation}
  \mathrm{Re}M(k)-a(k)I=-\mathcal{S}[\mathrm{Re}M(k)-a(k)I]\mathcal{S}^{-1}
\end{equation}
must be satisfied.
That is,
\begin{equation}
  \mathrm{Re}M(k)-a(k)I=A(k)\otimes\sigma_x+B(k)\otimes\sigma_y, \label{eq:ssls_app}
\end{equation}
where \(A\) and \(B\) are some \(2\times2\) matrices.
Assume that the jump operators act only on a unit cell and have translational symmetry,
\begin{equation}
  L_x=\sum_{s=A,B}\sum_{\omega=\alpha,\beta}l_{s,\omega}w_{x,s,\omega}. \label{eq:jump_app}
\end{equation}
Then \(M(k)\) does not have momentum dependence \(M(k)=\tilde{M}\), where \(\tilde{M}_{s\omega,s'\omega'}\coloneqq l_{s,\omega}l_{s',\omega'}^*\).
From these equations, Eq.\ \eqref{eq:ssls_app} is satisfied if \(\tilde{M}_{s\omega,s\omega}\) is constant and \(\tilde{M}_{s\alpha,s\beta}\) is purely imaginary.
These are equivalent to \(|l_{s,\omega}|=\gamma\) and \(\mathrm{arg}\,l_{s,\alpha}=\theta_s,~\mathrm{arg}\,l_{s,\beta}=\theta_s+(n_s-1/2)\pi~(\theta_s\in\mathbb{R},~n_s\in\mathbb{Z})\).
Inserting these equations into Eq.\ \eqref{eq:jump_app}, we get
\begin{align}
  L_x=&\gamma e^{i\theta_A}[w_{x,A,\alpha}+(-1)^{n_A+1}iw_{x,A,\beta}] \notag\\
  &+\gamma e^{i\theta_B}[w_{x,B,\alpha}+(-1)^{n_B+1}iw_{x,B,\beta}]. \label{eq:jump_sls_app}
\end{align}
Since \(c_{x,s}=(w_{x,s,\alpha}+iw_{x,s,\beta})/2\) and \(c_{x,s}^{\dagger}=(w_{x,s,\alpha}-iw_{x,s,\beta})/2\), Eq.\ \eqref{eq:jump_sls_app} implies that the jump operator contains only either the annihilation or creation operator of a fermion to retain shifted SLS.
For example, the jump operator in the main text in Eq.\ \eqref{eq:jump} contains only the annihilation operator of each fermion, and it satisfies the above relation.
Furthermore, the simultaneous one-body loss-gain process with equal strength described by the jump operators \(L_x=\gamma(c_{x,A}+c_{x,B}^{\dagger})\) satisfies the above relation, and the system can retain shifted SLS.
In contrast, the jump operator \(L_x=\gamma\sum_{s=A,B}w_{x,s,\alpha}=\gamma\sum_{s=A,B}(c_{x,s}+c_{x,s}^{\dagger})\) contains both the annihilation and creation operators of the fermion; thus, the system cannot retain shifted SLS.

In the case that \(\mathcal{S}\ne I_2\otimes\sigma_z\), we perform an unitary transformation such that \(U\mathcal{S}U^{\dagger}=I_2\otimes\sigma_z\), where \(U\) is the corresponding unitary operator of the transformation.
Then the matrix \(\tilde{M}\) is also transformed into \(\tilde{M}'=U\tilde{M}U^{\dagger}\).
The matrix element is given as \(\tilde{M}'_{s\omega,s'\omega'}={l'}_{s,\omega}{l'}_{s',\omega'}^*\), where \({l'}_{s,\omega}\coloneqq \sum_{s',\omega'}U_{s\omega,s'\omega'}l_{s',\omega'}\).
A similar derivation shows that the system retains shifted SLS if \(|{l'}_{s,\omega}|=\gamma\), \(\mathrm{arg}\,{l'}_{s,\alpha}=\theta'_s\), and \(\mathrm{arg}\,{l'}_{s,\beta}=\theta'_s+(n'_s-1/2)\pi\) (\(\theta'_s\in\mathbb{R}\) and \(n'_s\in\mathbb{Z}\)).
The original coefficients of the jump operators \(l_{s,\omega}\) are obtained as \(l_{s,\omega}=\sum_{s',\omega'}U^*_{s'\omega',s\omega}{l'}_{s',\omega'}\).

\section{Calculation of topological invariants}
\label{app:topoinv}
In this appendix, we derive the topological invariant in Eq.\ \eqref{eq:winding_ssh} by applying the method
in Ref.\ \cite{kawabata2019symmetry}.
First, we construct the vectors \(\ket{\phi_n^{\pm}}\) and \(\bket{\phi_n^{\pm}}\) given in Eq.\ \eqref{eq:subsector}.
Let \(\ket{\psi_{n}}\) and \(\bket{\psi_{n}}\) be a right and left biorthonormal eigenvectors of \(Z'(k)\),
\begin{equation}
  Z'(k)\ket{\psi_{n}}=\lambda_{n}\ket{\psi_{n}},~Z'^{\dagger}(k)\bket{\psi_{n}}=\lambda_{n}^*\bket{\psi_{n}}, \label{appeq:eigvec}
\end{equation}
respectively.
We suppose that \(Z'\) has pseudo-Hermiticity \(Z'(k)=\eta Z'^{\dagger}(k)\eta^{-1}\); then the right and left eigenvectors are connected as
\begin{equation}
  \eta\bket{\psi_{n}}=\sum_m A_{mn}\ket{\psi_m} \label{appeq:eigvec_eta}
\end{equation}
holds.
Since \(A=[A_{mn}]\) is Hermitian, \(A\) is diagonalized by a unitary matrix \(G\),
\begin{equation}
  G^{\dagger}A G=\mathrm{diag}\{\xi_1,\xi_2,\cdots,\xi_N\}, \label{appeq:diagonalize_c}
\end{equation}
where \(\xi_i\in\mathbb{R}\).
By using the matrix \(G\), biorthonormal vectors \(\ket{\phi_n^{\pm}}\) and \(\bket{\phi_n^{\pm}}\) are constructed as
\begin{align}
  \ket{\phi_n}&\coloneqq \sum_m\sqrt{|\xi_n|}G_{mn}\ket{\psi_m}, \label{appeq:phi_right} \\
  \bket{\phi_n}&\coloneqq \sum_m\frac{G_{mn}}{\sqrt{|\xi_n|}}\bket{\psi_m}. \label{appeq:phi_left}
\end{align}
Then the state space is divided into two subsectors, one of which is spanned by states with
\begin{equation}
  \eta\ket{\phi_n}=\bket{\phi_n}
\end{equation}
and the other is spanned by states with
\begin{equation}
  \eta\ket{\phi_n}=-\bket{\phi_n}.
\end{equation}
We denote the former as \(\ket{\phi_n^+}~(\bket{\phi_n^+})\), and the latter as \(\ket{\phi_n^-}~(\bket{\phi_n^-})\).

The eigenvectors of \(Z'(k)\) are given by
\begin{align}
  &\ket{\psi_{p,q}(k)}=\frac{1}{2}\begin{pmatrix}
  \frac{(-1)^{q}}{{\lambda'_p}}[(-1)^{p+1}t_k^*+i\gamma^2] \\
  (-1)^{p+1}i  \\
  \frac{(-1)^{q}}{{\lambda'_p}}[it_k^*+(-1)^{p}\gamma^2] \\
  1
  \end{pmatrix},\label{appeq:eig_right}\\
  &\bket{\psi_{p,q}(k)}=\frac{1}{2}\begin{pmatrix}
  \frac{(-1)^{q}}{{\lambda'_p}^*}[(-1)^{p+1}t_k^*-i\gamma^2] \\
  (-1)^{p+1}i \\
  \frac{(-1)^{q}}{{\lambda'_p}^*}[it_k^*+(-1)^{p+1}\gamma^2] \\
  1
  \end{pmatrix},\label{appeq:eig_left} \\
  &{\lambda'_p}\coloneqq\sqrt{|t_k|^2-\gamma^4+(-1)^{p+1}2i\gamma^2\mathrm{Re}t_k}.
\end{align}
Then the vectors \(\ket{\phi_n^{\pm}(k)}\) and \(\bket{\phi_n^{\pm}(k)}\) \((n\in\{0,1\})\) are calculated as
\begin{align}
  \ket{\phi_n^+(k)}&=\frac{1}{\sqrt{2}}[\ket{\psi_{0,n}(k)}+i\ket{\psi_{1,n}(k)}], \\
  \bket{\phi_n^+(k)}&=\frac{1}{\sqrt{2}}[\bket{\psi_{0,n}(k)}+i\bket{\psi_{1,n}(k)}], \\
  \ket{\phi_n^-(k)}&=\frac{1}{\sqrt{2}}[-\ket{\psi_{0,n}(k)}+i\ket{\psi_{1,n}(k)}], \\
  \bket{\phi_n^-(k)}&=\frac{1}{\sqrt{2}}[-\bket{\psi_{0,n}(k)}+i\bket{\psi_{1,n}(k)}]. \label{appeq:phi_n_minus_left}.
\end{align}
By using the projection operators in plus and minus subsectors,
\begin{equation}
  P^{\pm}(k)\coloneqq \sum_n \ket{\phi_n^{\pm}(k)}\bbra{\phi_n^{\pm}(k)}.
\end{equation}
Each subsection of the matrix \(Z'\) is provided as
\begin{equation}
  Z'_{\pm}(k)\coloneqq P^{\pm}(k)Z'(k)P^{\pm}(k),
\end{equation}
and their explicit expressions are given as
\begin{align}
  Z'_{\pm}(k)=\pm&\frac{\lambda_{0,0}(k)+\lambda_{1,0}(k)}{2}\nonumber\\
  &\times(\ket{\phi_0^{\pm}(k)}\bbra{\phi_0^{\pm}(k)}-\ket{\phi_1^{\pm}(k)}\bbra{\phi_1^{\pm}(k)}),
\end{align}
where $\lambda_{p,q}(k)$ is the eigenvalue of \(Z\) given in Eq.\ \eqref{eq:eigval}.
\(Z'_{\pm}\) has SLS \(Z'_{\pm}(k)=-\mathcal{S}_{\pm}Z'_{\pm}(k)\mathcal{S}_{\pm}^{-1},~\mathcal{S}_{\pm}=I_2\otimes\sigma_z\), thus we can define the winding numbers in Eq.\ \eqref{eq:topoinv_winding} for each subsector.
We have
\begin{align}
  &\mathrm{tr}\left[\mathcal{S}_{\pm}Z_{\pm}^{'-1}(k)\frac{dZ'_{\pm}(k)}{dk}\right]\nonumber\\
  &=-2\left(\bbraket{\phi_1^{\pm}(k)|\frac{d\phi_0^{\pm}(k)}{dk}}-\bbraket{\phi_0^{\pm}(k)|\frac{d\phi_1^{\pm}(k)}{dk}}\right)
\end{align}
After some algebra using Eqs.\ \eqref{appeq:eig_right}-\eqref{appeq:phi_n_minus_left}, we get
\begin{equation}
  \begin{split}
    &\mathrm{tr}\left[\mathcal{S}_{+}Z_{+}^{'-1}(k)\frac{dZ'_{+}(k)}{dk}\right]=\mathrm{tr}\left[\mathcal{S}_{-}Z_{-}^{'-1}(k)\frac{dZ'_{-}(k)}{dk}\right] \\
    &=\frac{d}{dk}\left[\ln(t_k^*-i\gamma^2)+\ln(t_k^*-i\gamma^2)-\ln{\lambda'_0}-\ln{\lambda'_0}^*\right].
  \end{split}
\end{equation}
In our model \({\lambda'_0}\) does not have a winding number; thus, \(\displaystyle\frac{1}{2\pi i}\int_{\mathrm{BZ}}dk\frac{d}{dk}\ln{\lambda'_0}=\frac{1}{2\pi i}\int_{\mathrm{BZ}}dk\frac{d}{dk}\ln{\lambda'_0}^*=0\) holds.
Then we have \(w_+=w_-\), and they are given as
\begin{equation}
  w_{\pm}=\frac{1}{4\pi i}\int_{\mathrm{BZ}}dk\left[\frac{d}{dk}\ln(t^*_k-i\gamma^2)+\frac{d}{dk}\ln(t^*_k+i\gamma^2)\right]. \label{eq:winding_ssh_ap}
\end{equation}
As a result, the topological invariant for \(Z\) is given as \(w=w_++w_-\) and is provided as Eq.\ \eqref{eq:winding_ssh}.

\bibliographystyle{apsrev4-2}
\bibliography{ssh}

\begin{thebibliography}{59}%
\makeatletter
\providecommand \@ifxundefined [1]{%
 \@ifx{#1\undefined}
}%
\providecommand \@ifnum [1]{%
 \ifnum #1\expandafter \@firstoftwo
 \else \expandafter \@secondoftwo
 \fi
}%
\providecommand \@ifx [1]{%
 \ifx #1\expandafter \@firstoftwo
 \else \expandafter \@secondoftwo
 \fi
}%
\providecommand \natexlab [1]{#1}%
\providecommand \enquote  [1]{``#1''}%
\providecommand \bibnamefont  [1]{#1}%
\providecommand \bibfnamefont [1]{#1}%
\providecommand \citenamefont [1]{#1}%
\providecommand \href@noop [0]{\@secondoftwo}%
\providecommand \href [0]{\begingroup \@sanitize@url \@href}%
\providecommand \@href[1]{\@@startlink{#1}\@@href}%
\providecommand \@@href[1]{\endgroup#1\@@endlink}%
\providecommand \@sanitize@url [0]{\catcode `\\12\catcode `\$12\catcode
  `\&12\catcode `\#12\catcode `\^12\catcode `\_12\catcode `\%12\relax}%
\providecommand \@@startlink[1]{}%
\providecommand \@@endlink[0]{}%
\providecommand \url  [0]{\begingroup\@sanitize@url \@url }%
\providecommand \@url [1]{\endgroup\@href {#1}{\urlprefix }}%
\providecommand \urlprefix  [0]{URL }%
\providecommand \Eprint [0]{\href }%
\providecommand \doibase [0]{https://doi.org/}%
\providecommand \selectlanguage [0]{\@gobble}%
\providecommand \bibinfo  [0]{\@secondoftwo}%
\providecommand \bibfield  [0]{\@secondoftwo}%
\providecommand \translation [1]{[#1]}%
\providecommand \BibitemOpen [0]{}%
\providecommand \bibitemStop [0]{}%
\providecommand \bibitemNoStop [0]{.\EOS\space}%
\providecommand \EOS [0]{\spacefactor3000\relax}%
\providecommand \BibitemShut  [1]{\csname bibitem#1\endcsname}%
\let\auto@bib@innerbib\@empty
\bibitem [{\citenamefont {Esaki}\ \emph {et~al.}(2011)\citenamefont {Esaki},
  \citenamefont {Sato}, \citenamefont {Hasebe},\ and\ \citenamefont
  {Kohmoto}}]{esaki2011edge}%
  \BibitemOpen
  \bibfield  {author} {\bibinfo {author} {\bibfnamefont {K.}~\bibnamefont
  {Esaki}}, \bibinfo {author} {\bibfnamefont {M.}~\bibnamefont {Sato}},
  \bibinfo {author} {\bibfnamefont {K.}~\bibnamefont {Hasebe}},\ and\ \bibinfo
  {author} {\bibfnamefont {M.}~\bibnamefont {Kohmoto}},\ }\href
  {https://link.aps.org/doi/10.1103/PhysRevB.84.205128} {\bibfield  {journal}
  {\bibinfo  {journal} {Phys. Rev. B}\ }\textbf {\bibinfo {volume} {84}},\
  \bibinfo {pages} {205128} (\bibinfo {year} {2011})}\BibitemShut {NoStop}%
\bibitem [{\citenamefont {Lee}(2016)}]{lee2016anomalous}%
  \BibitemOpen
  \bibfield  {author} {\bibinfo {author} {\bibfnamefont {T.~E.}\ \bibnamefont
  {Lee}},\ }\href {https://link.aps.org/doi/10.1103/PhysRevLett.116.133903}
  {\bibfield  {journal} {\bibinfo  {journal} {Phys. Rev. Lett.}\ }\textbf
  {\bibinfo {volume} {116}},\ \bibinfo {pages} {133903} (\bibinfo {year}
  {2016})}\BibitemShut {NoStop}%
\bibitem [{\citenamefont {Lieu}(2018)}]{lieu2018topological}%
  \BibitemOpen
  \bibfield  {author} {\bibinfo {author} {\bibfnamefont {S.}~\bibnamefont
  {Lieu}},\ }\href {https://link.aps.org/doi/10.1103/PhysRevB.97.045106}
  {\bibfield  {journal} {\bibinfo  {journal} {Phys. Rev. B}\ }\textbf {\bibinfo
  {volume} {97}},\ \bibinfo {pages} {045106} (\bibinfo {year}
  {2018})}\BibitemShut {NoStop}%
\bibitem [{\citenamefont {Kawabata}\ \emph {et~al.}(2018)\citenamefont
  {Kawabata}, \citenamefont {Ashida}, \citenamefont {Katsura},\ and\
  \citenamefont {Ueda}}]{kawabata2018parity}%
  \BibitemOpen
  \bibfield  {author} {\bibinfo {author} {\bibfnamefont {K.}~\bibnamefont
  {Kawabata}}, \bibinfo {author} {\bibfnamefont {Y.}~\bibnamefont {Ashida}},
  \bibinfo {author} {\bibfnamefont {H.}~\bibnamefont {Katsura}},\ and\ \bibinfo
  {author} {\bibfnamefont {M.}~\bibnamefont {Ueda}},\ }\href
  {https://link.aps.org/doi/10.1103/PhysRevB.98.085116} {\bibfield  {journal}
  {\bibinfo  {journal} {Phys. Rev. B}\ }\textbf {\bibinfo {volume} {98}},\
  \bibinfo {pages} {085116} (\bibinfo {year} {2018})}\BibitemShut {NoStop}%
\bibitem [{\citenamefont {Yokomizo}\ and\ \citenamefont
  {Murakami}(2019)}]{yokomizo2019non}%
  \BibitemOpen
  \bibfield  {author} {\bibinfo {author} {\bibfnamefont {K.}~\bibnamefont
  {Yokomizo}}\ and\ \bibinfo {author} {\bibfnamefont {S.}~\bibnamefont
  {Murakami}},\ }\href
  {https://link.aps.org/doi/10.1103/PhysRevLett.123.066404} {\bibfield
  {journal} {\bibinfo  {journal} {Phys. Rev. Lett.}\ }\textbf {\bibinfo
  {volume} {123}},\ \bibinfo {pages} {066404} (\bibinfo {year}
  {2019})}\BibitemShut {NoStop}%
\bibitem [{\citenamefont {Liu}\ \emph {et~al.}(2019)\citenamefont {Liu},
  \citenamefont {Zhang}, \citenamefont {Ai}, \citenamefont {Gong},
  \citenamefont {Kawabata}, \citenamefont {Ueda},\ and\ \citenamefont
  {Nori}}]{liu2019second}%
  \BibitemOpen
  \bibfield  {author} {\bibinfo {author} {\bibfnamefont {T.}~\bibnamefont
  {Liu}}, \bibinfo {author} {\bibfnamefont {Y.-R.}\ \bibnamefont {Zhang}},
  \bibinfo {author} {\bibfnamefont {Q.}~\bibnamefont {Ai}}, \bibinfo {author}
  {\bibfnamefont {Z.}~\bibnamefont {Gong}}, \bibinfo {author} {\bibfnamefont
  {K.}~\bibnamefont {Kawabata}}, \bibinfo {author} {\bibfnamefont
  {M.}~\bibnamefont {Ueda}},\ and\ \bibinfo {author} {\bibfnamefont
  {F.}~\bibnamefont {Nori}},\ }\href
  {https://link.aps.org/doi/10.1103/PhysRevLett.122.076801} {\bibfield
  {journal} {\bibinfo  {journal} {Phys. Rev. Lett.}\ }\textbf {\bibinfo
  {volume} {122}},\ \bibinfo {pages} {076801} (\bibinfo {year}
  {2019})}\BibitemShut {NoStop}%
\bibitem [{\citenamefont {Zeng}\ \emph {et~al.}(2020)\citenamefont {Zeng},
  \citenamefont {Yang},\ and\ \citenamefont {Xu}}]{zeng2020topological}%
  \BibitemOpen
  \bibfield  {author} {\bibinfo {author} {\bibfnamefont {Q.-B.}\ \bibnamefont
  {Zeng}}, \bibinfo {author} {\bibfnamefont {Y.-B.}\ \bibnamefont {Yang}},\
  and\ \bibinfo {author} {\bibfnamefont {Y.}~\bibnamefont {Xu}},\ }\href
  {https://link.aps.org/doi/10.1103/PhysRevB.101.020201} {\bibfield  {journal}
  {\bibinfo  {journal} {Phys. Rev. B}\ }\textbf {\bibinfo {volume} {101}},\
  \bibinfo {pages} {020201(R)} (\bibinfo {year} {2020})}\BibitemShut {NoStop}%
\bibitem [{\citenamefont {Ashida}\ \emph {et~al.}(2020)\citenamefont {Ashida},
  \citenamefont {Gong},\ and\ \citenamefont {Ueda}}]{ashida2020non}%
  \BibitemOpen
  \bibfield  {author} {\bibinfo {author} {\bibfnamefont {Y.}~\bibnamefont
  {Ashida}}, \bibinfo {author} {\bibfnamefont {Z.}~\bibnamefont {Gong}},\ and\
  \bibinfo {author} {\bibfnamefont {M.}~\bibnamefont {Ueda}},\ }\href
  {https://doi.org/10.1080/00018732.2021.1876991} {\bibfield  {journal}
  {\bibinfo  {journal} {Adv. Phys.}\ }\textbf {\bibinfo {volume} {69}},\
  \bibinfo {pages} {249} (\bibinfo {year} {2020})}\BibitemShut {NoStop}%
\bibitem [{\citenamefont {Bergholtz}\ \emph {et~al.}(2021)\citenamefont
  {Bergholtz}, \citenamefont {Budich},\ and\ \citenamefont
  {Kunst}}]{bergholtz2021exceptional}%
  \BibitemOpen
  \bibfield  {author} {\bibinfo {author} {\bibfnamefont {E.~J.}\ \bibnamefont
  {Bergholtz}}, \bibinfo {author} {\bibfnamefont {J.~C.}\ \bibnamefont
  {Budich}},\ and\ \bibinfo {author} {\bibfnamefont {F.~K.}\ \bibnamefont
  {Kunst}},\ }\href {https://link.aps.org/doi/10.1103/RevModPhys.93.015005}
  {\bibfield  {journal} {\bibinfo  {journal} {Rev. Mod. Phys.}\ }\textbf
  {\bibinfo {volume} {93}},\ \bibinfo {pages} {015005} (\bibinfo {year}
  {2021})}\BibitemShut {NoStop}%
\bibitem [{\citenamefont {Gong}\ \emph {et~al.}(2018)\citenamefont {Gong},
  \citenamefont {Ashida}, \citenamefont {Kawabata}, \citenamefont {Takasan},
  \citenamefont {Higashikawa},\ and\ \citenamefont
  {Ueda}}]{gong2018topological}%
  \BibitemOpen
  \bibfield  {author} {\bibinfo {author} {\bibfnamefont {Z.}~\bibnamefont
  {Gong}}, \bibinfo {author} {\bibfnamefont {Y.}~\bibnamefont {Ashida}},
  \bibinfo {author} {\bibfnamefont {K.}~\bibnamefont {Kawabata}}, \bibinfo
  {author} {\bibfnamefont {K.}~\bibnamefont {Takasan}}, \bibinfo {author}
  {\bibfnamefont {S.}~\bibnamefont {Higashikawa}},\ and\ \bibinfo {author}
  {\bibfnamefont {M.}~\bibnamefont {Ueda}},\ }\href
  {https://link.aps.org/doi/10.1103/PhysRevX.8.031079} {\bibfield  {journal}
  {\bibinfo  {journal} {Phys. Rev. X}\ }\textbf {\bibinfo {volume} {8}},\
  \bibinfo {pages} {031079} (\bibinfo {year} {2018})}\BibitemShut {NoStop}%
\bibitem [{\citenamefont {Kawabata}\ \emph
  {et~al.}(2019{\natexlab{a}})\citenamefont {Kawabata}, \citenamefont
  {Higashikawa}, \citenamefont {Gong}, \citenamefont {Ashida},\ and\
  \citenamefont {Ueda}}]{kawabata2019topological}%
  \BibitemOpen
  \bibfield  {author} {\bibinfo {author} {\bibfnamefont {K.}~\bibnamefont
  {Kawabata}}, \bibinfo {author} {\bibfnamefont {S.}~\bibnamefont
  {Higashikawa}}, \bibinfo {author} {\bibfnamefont {Z.}~\bibnamefont {Gong}},
  \bibinfo {author} {\bibfnamefont {Y.}~\bibnamefont {Ashida}},\ and\ \bibinfo
  {author} {\bibfnamefont {M.}~\bibnamefont {Ueda}},\ }\href
  {https://doi.org/10.1038/s41467-018-08254-y} {\bibfield  {journal} {\bibinfo
  {journal} {Nat. Commun.}\ }\textbf {\bibinfo {volume} {10}},\ \bibinfo
  {pages} {1} (\bibinfo {year} {2019}{\natexlab{a}})}\BibitemShut {NoStop}%
\bibitem [{\citenamefont {Kawabata}\ \emph
  {et~al.}(2019{\natexlab{b}})\citenamefont {Kawabata}, \citenamefont
  {Shiozaki}, \citenamefont {Ueda},\ and\ \citenamefont
  {Sato}}]{kawabata2019symmetry}%
  \BibitemOpen
  \bibfield  {author} {\bibinfo {author} {\bibfnamefont {K.}~\bibnamefont
  {Kawabata}}, \bibinfo {author} {\bibfnamefont {K.}~\bibnamefont {Shiozaki}},
  \bibinfo {author} {\bibfnamefont {M.}~\bibnamefont {Ueda}},\ and\ \bibinfo
  {author} {\bibfnamefont {M.}~\bibnamefont {Sato}},\ }\href
  {https://link.aps.org/doi/10.1103/PhysRevX.9.041015} {\bibfield  {journal}
  {\bibinfo  {journal} {Phys. Rev. X}\ }\textbf {\bibinfo {volume} {9}},\
  \bibinfo {pages} {041015} (\bibinfo {year} {2019}{\natexlab{b}})}\BibitemShut
  {NoStop}%
\bibitem [{\citenamefont {Yao}\ and\ \citenamefont {Wang}(2018)}]{yao2018edge}%
  \BibitemOpen
  \bibfield  {author} {\bibinfo {author} {\bibfnamefont {S.}~\bibnamefont
  {Yao}}\ and\ \bibinfo {author} {\bibfnamefont {Z.}~\bibnamefont {Wang}},\
  }\href {https://link.aps.org/doi/10.1103/PhysRevLett.121.086803} {\bibfield
  {journal} {\bibinfo  {journal} {Phys. Rev. Lett.}\ }\textbf {\bibinfo
  {volume} {121}},\ \bibinfo {pages} {086803} (\bibinfo {year}
  {2018})}\BibitemShut {NoStop}%
\bibitem [{\citenamefont {Okuma}\ \emph {et~al.}(2020)\citenamefont {Okuma},
  \citenamefont {Kawabata}, \citenamefont {Shiozaki},\ and\ \citenamefont
  {Sato}}]{okuma2020topological}%
  \BibitemOpen
  \bibfield  {author} {\bibinfo {author} {\bibfnamefont {N.}~\bibnamefont
  {Okuma}}, \bibinfo {author} {\bibfnamefont {K.}~\bibnamefont {Kawabata}},
  \bibinfo {author} {\bibfnamefont {K.}~\bibnamefont {Shiozaki}},\ and\
  \bibinfo {author} {\bibfnamefont {M.}~\bibnamefont {Sato}},\ }\href
  {https://link.aps.org/doi/10.1103/PhysRevLett.124.086801} {\bibfield
  {journal} {\bibinfo  {journal} {Phys. Rev. Lett.}\ }\textbf {\bibinfo
  {volume} {124}},\ \bibinfo {pages} {086801} (\bibinfo {year}
  {2020})}\BibitemShut {NoStop}%
\bibitem [{\citenamefont {Kawasaki}\ \emph {et~al.}(2020)\citenamefont
  {Kawasaki}, \citenamefont {Mochizuki}, \citenamefont {Kawakami},\ and\
  \citenamefont {Obuse}}]{kawasaki2020bulk}%
  \BibitemOpen
  \bibfield  {author} {\bibinfo {author} {\bibfnamefont {M.}~\bibnamefont
  {Kawasaki}}, \bibinfo {author} {\bibfnamefont {K.}~\bibnamefont {Mochizuki}},
  \bibinfo {author} {\bibfnamefont {N.}~\bibnamefont {Kawakami}},\ and\
  \bibinfo {author} {\bibfnamefont {H.}~\bibnamefont {Obuse}},\ }\href
  {https://doi.org/10.1093/ptep/ptaa034} {\bibfield  {journal} {\bibinfo
  {journal} {Prog. Theor. Exp. Phys.}\ }\textbf {\bibinfo {volume} {2020}},\
  \bibinfo {pages} {12A105} (\bibinfo {year} {2020})}\BibitemShut {NoStop}%
\bibitem [{\citenamefont {Zhang}\ \emph {et~al.}(2020)\citenamefont {Zhang},
  \citenamefont {Yang},\ and\ \citenamefont {Fang}}]{zhang2020correspondence}%
  \BibitemOpen
  \bibfield  {author} {\bibinfo {author} {\bibfnamefont {K.}~\bibnamefont
  {Zhang}}, \bibinfo {author} {\bibfnamefont {Z.}~\bibnamefont {Yang}},\ and\
  \bibinfo {author} {\bibfnamefont {C.}~\bibnamefont {Fang}},\ }\href
  {https://link.aps.org/doi/10.1103/PhysRevLett.125.126402} {\bibfield
  {journal} {\bibinfo  {journal} {Phys. Rev. Lett.}\ }\textbf {\bibinfo
  {volume} {125}},\ \bibinfo {pages} {126402} (\bibinfo {year}
  {2020})}\BibitemShut {NoStop}%
\bibitem [{\citenamefont {Sone}\ \emph {et~al.}(2020)\citenamefont {Sone},
  \citenamefont {Ashida},\ and\ \citenamefont {Sagawa}}]{sone2020exceptional}%
  \BibitemOpen
  \bibfield  {author} {\bibinfo {author} {\bibfnamefont {K.}~\bibnamefont
  {Sone}}, \bibinfo {author} {\bibfnamefont {Y.}~\bibnamefont {Ashida}},\ and\
  \bibinfo {author} {\bibfnamefont {T.}~\bibnamefont {Sagawa}},\ }\href
  {https://doi.org/10.1038/s41467-020-19488-0} {\bibfield  {journal} {\bibinfo
  {journal} {Nat. Commun.}\ }\textbf {\bibinfo {volume} {11}},\ \bibinfo
  {pages} {1} (\bibinfo {year} {2020})}\BibitemShut {NoStop}%
\bibitem [{\citenamefont {Borgnia}\ \emph {et~al.}(2020)\citenamefont
  {Borgnia}, \citenamefont {Kruchkov},\ and\ \citenamefont
  {Slager}}]{borgnia2020non}%
  \BibitemOpen
  \bibfield  {author} {\bibinfo {author} {\bibfnamefont {D.~S.}\ \bibnamefont
  {Borgnia}}, \bibinfo {author} {\bibfnamefont {A.~J.}\ \bibnamefont
  {Kruchkov}},\ and\ \bibinfo {author} {\bibfnamefont {R.-J.}\ \bibnamefont
  {Slager}},\ }\href {https://link.aps.org/doi/10.1103/PhysRevLett.124.056802}
  {\bibfield  {journal} {\bibinfo  {journal} {Physical review letters}\
  }\textbf {\bibinfo {volume} {124}},\ \bibinfo {pages} {056802} (\bibinfo
  {year} {2020})}\BibitemShut {NoStop}%
\bibitem [{\citenamefont {Weimann}\ \emph {et~al.}(2017)\citenamefont
  {Weimann}, \citenamefont {Kremer}, \citenamefont {Plotnik}, \citenamefont
  {Lumer}, \citenamefont {Nolte}, \citenamefont {Makris}, \citenamefont
  {Segev}, \citenamefont {Rechtsman},\ and\ \citenamefont
  {Szameit}}]{weimann2017topologically}%
  \BibitemOpen
  \bibfield  {author} {\bibinfo {author} {\bibfnamefont {S.}~\bibnamefont
  {Weimann}}, \bibinfo {author} {\bibfnamefont {M.}~\bibnamefont {Kremer}},
  \bibinfo {author} {\bibfnamefont {Y.}~\bibnamefont {Plotnik}}, \bibinfo
  {author} {\bibfnamefont {Y.}~\bibnamefont {Lumer}}, \bibinfo {author}
  {\bibfnamefont {S.}~\bibnamefont {Nolte}}, \bibinfo {author} {\bibfnamefont
  {K.~G.}\ \bibnamefont {Makris}}, \bibinfo {author} {\bibfnamefont
  {M.}~\bibnamefont {Segev}}, \bibinfo {author} {\bibfnamefont {M.~C.}\
  \bibnamefont {Rechtsman}},\ and\ \bibinfo {author} {\bibfnamefont
  {A.}~\bibnamefont {Szameit}},\ }\href {https://doi.org/10.1038/nmat4811}
  {\bibfield  {journal} {\bibinfo  {journal} {Nat. Mater.}\ }\textbf {\bibinfo
  {volume} {16}},\ \bibinfo {pages} {433} (\bibinfo {year} {2017})}\BibitemShut
  {NoStop}%
\bibitem [{\citenamefont {Bandres}\ \emph {et~al.}(2018)\citenamefont
  {Bandres}, \citenamefont {Wittek}, \citenamefont {Harari}, \citenamefont
  {Parto}, \citenamefont {Ren}, \citenamefont {Segev}, \citenamefont
  {Christodoulides},\ and\ \citenamefont
  {Khajavikhan}}]{bandres2018topological}%
  \BibitemOpen
  \bibfield  {author} {\bibinfo {author} {\bibfnamefont {M.~A.}\ \bibnamefont
  {Bandres}}, \bibinfo {author} {\bibfnamefont {S.}~\bibnamefont {Wittek}},
  \bibinfo {author} {\bibfnamefont {G.}~\bibnamefont {Harari}}, \bibinfo
  {author} {\bibfnamefont {M.}~\bibnamefont {Parto}}, \bibinfo {author}
  {\bibfnamefont {J.}~\bibnamefont {Ren}}, \bibinfo {author} {\bibfnamefont
  {M.}~\bibnamefont {Segev}}, \bibinfo {author} {\bibfnamefont {D.~N.}\
  \bibnamefont {Christodoulides}},\ and\ \bibinfo {author} {\bibfnamefont
  {M.}~\bibnamefont {Khajavikhan}},\ }\href
  {https://doi.org/10.1126/science.aar4005} {\bibfield  {journal} {\bibinfo
  {journal} {Science}\ }\textbf {\bibinfo {volume} {359}} (\bibinfo {year}
  {2018})}\BibitemShut {NoStop}%
\bibitem [{\citenamefont {Mochizuki}\ \emph {et~al.}(2016)\citenamefont
  {Mochizuki}, \citenamefont {Kim},\ and\ \citenamefont
  {Obuse}}]{mochizuki2016}%
  \BibitemOpen
  \bibfield  {author} {\bibinfo {author} {\bibfnamefont {K.}~\bibnamefont
  {Mochizuki}}, \bibinfo {author} {\bibfnamefont {D.}~\bibnamefont {Kim}},\
  and\ \bibinfo {author} {\bibfnamefont {H.}~\bibnamefont {Obuse}},\ }\href
  {https://link.aps.org/doi/10.1103/PhysRevA.93.062116} {\bibfield  {journal}
  {\bibinfo  {journal} {Phys. Rev. A}\ }\textbf {\bibinfo {volume} {93}},\
  \bibinfo {pages} {062116} (\bibinfo {year} {2016})}\BibitemShut {NoStop}%
\bibitem [{\citenamefont {Xiao}\ \emph {et~al.}(2017)\citenamefont {Xiao},
  \citenamefont {Zhan}, \citenamefont {Bian}, \citenamefont {Wang},
  \citenamefont {Zhang}, \citenamefont {Wang}, \citenamefont {Li},
  \citenamefont {Mochizuki}, \citenamefont {Kim}, \citenamefont {Kawakami},
  \citenamefont {Yi}, \citenamefont {Obuse}, \citenamefont {Sanders},\ and\
  \citenamefont {Xue}}]{xiao2017observation}%
  \BibitemOpen
  \bibfield  {author} {\bibinfo {author} {\bibfnamefont {L.}~\bibnamefont
  {Xiao}}, \bibinfo {author} {\bibfnamefont {X.}~\bibnamefont {Zhan}}, \bibinfo
  {author} {\bibfnamefont {Z.~H.}\ \bibnamefont {Bian}}, \bibinfo {author}
  {\bibfnamefont {K.~K.}\ \bibnamefont {Wang}}, \bibinfo {author}
  {\bibfnamefont {X.}~\bibnamefont {Zhang}}, \bibinfo {author} {\bibfnamefont
  {X.~P.}\ \bibnamefont {Wang}}, \bibinfo {author} {\bibfnamefont
  {J.}~\bibnamefont {Li}}, \bibinfo {author} {\bibfnamefont {K.}~\bibnamefont
  {Mochizuki}}, \bibinfo {author} {\bibfnamefont {D.}~\bibnamefont {Kim}},
  \bibinfo {author} {\bibfnamefont {N.}~\bibnamefont {Kawakami}}, \bibinfo
  {author} {\bibfnamefont {W.}~\bibnamefont {Yi}}, \bibinfo {author}
  {\bibfnamefont {H.}~\bibnamefont {Obuse}}, \bibinfo {author} {\bibfnamefont
  {B.~C.}\ \bibnamefont {Sanders}},\ and\ \bibinfo {author} {\bibfnamefont
  {P.}~\bibnamefont {Xue}},\ }\href {https://doi.org/10.1038/nphys4204}
  {\bibfield  {journal} {\bibinfo  {journal} {Nat. Phys.}\ }\textbf {\bibinfo
  {volume} {13}},\ \bibinfo {pages} {1117} (\bibinfo {year}
  {2017})}\BibitemShut {NoStop}%
\bibitem [{\citenamefont {Xiao}\ \emph {et~al.}(2020)\citenamefont {Xiao},
  \citenamefont {Deng}, \citenamefont {Wang}, \citenamefont {Zhu},
  \citenamefont {Wang}, \citenamefont {Yi},\ and\ \citenamefont
  {Xue}}]{xiao2020non}%
  \BibitemOpen
  \bibfield  {author} {\bibinfo {author} {\bibfnamefont {L.}~\bibnamefont
  {Xiao}}, \bibinfo {author} {\bibfnamefont {T.}~\bibnamefont {Deng}}, \bibinfo
  {author} {\bibfnamefont {K.}~\bibnamefont {Wang}}, \bibinfo {author}
  {\bibfnamefont {G.}~\bibnamefont {Zhu}}, \bibinfo {author} {\bibfnamefont
  {Z.}~\bibnamefont {Wang}}, \bibinfo {author} {\bibfnamefont {W.}~\bibnamefont
  {Yi}},\ and\ \bibinfo {author} {\bibfnamefont {P.}~\bibnamefont {Xue}},\
  }\href {https://doi.org/10.1038/s41567-020-0836-6} {\bibfield  {journal}
  {\bibinfo  {journal} {Nature Physics}\ }\textbf {\bibinfo {volume} {16}},\
  \bibinfo {pages} {761} (\bibinfo {year} {2020})}\BibitemShut {NoStop}%
\bibitem [{\citenamefont {Breuer}\ and\ \citenamefont
  {Petruccione}(2002)}]{breuer2002theory}%
  \BibitemOpen
  \bibfield  {author} {\bibinfo {author} {\bibfnamefont {H.-P.}\ \bibnamefont
  {Breuer}}\ and\ \bibinfo {author} {\bibfnamefont {F.}~\bibnamefont
  {Petruccione}},\ }\href@noop {} {\emph {\bibinfo {title} {The theory of open
  quantum systems}}}\ (\bibinfo  {publisher} {Oxford University Press on
  Demand},\ \bibinfo {year} {2002})\BibitemShut {NoStop}%
\bibitem [{\citenamefont {Landi}\ \emph {et~al.}(2021)\citenamefont {Landi},
  \citenamefont {Poletti},\ and\ \citenamefont {Schaller}}]{landi2021non}%
  \BibitemOpen
  \bibfield  {author} {\bibinfo {author} {\bibfnamefont {G.~T.}\ \bibnamefont
  {Landi}}, \bibinfo {author} {\bibfnamefont {D.}~\bibnamefont {Poletti}},\
  and\ \bibinfo {author} {\bibfnamefont {G.}~\bibnamefont {Schaller}},\
  }\href@noop {} {\bibfield  {journal} {\bibinfo  {journal} {arXiv preprint
  arXiv:2104.14350}\ } (\bibinfo {year} {2021})}\BibitemShut {NoStop}%
\bibitem [{\citenamefont {Prosen}(2011)}]{prosen2011open}%
  \BibitemOpen
  \bibfield  {author} {\bibinfo {author} {\bibfnamefont {T.}~\bibnamefont
  {Prosen}},\ }\href {https://link.aps.org/doi/10.1103/PhysRevLett.106.217206}
  {\bibfield  {journal} {\bibinfo  {journal} {Phys. Rev. Lett.}\ }\textbf
  {\bibinfo {volume} {106}},\ \bibinfo {pages} {217206} (\bibinfo {year}
  {2011})}\BibitemShut {NoStop}%
\bibitem [{\citenamefont {Minganti}\ \emph {et~al.}(2018)\citenamefont
  {Minganti}, \citenamefont {Biella}, \citenamefont {Bartolo},\ and\
  \citenamefont {Ciuti}}]{minganti2018spectral}%
  \BibitemOpen
  \bibfield  {author} {\bibinfo {author} {\bibfnamefont {F.}~\bibnamefont
  {Minganti}}, \bibinfo {author} {\bibfnamefont {A.}~\bibnamefont {Biella}},
  \bibinfo {author} {\bibfnamefont {N.}~\bibnamefont {Bartolo}},\ and\ \bibinfo
  {author} {\bibfnamefont {C.}~\bibnamefont {Ciuti}},\ }\href
  {https://link.aps.org/doi/10.1103/PhysRevA.98.042118} {\bibfield  {journal}
  {\bibinfo  {journal} {Phys. Rev. A}\ }\textbf {\bibinfo {volume} {98}},\
  \bibinfo {pages} {042118} (\bibinfo {year} {2018})}\BibitemShut {NoStop}%
\bibitem [{\citenamefont {Mori}\ and\ \citenamefont
  {Shirai}(2020)}]{mori2020resolving}%
  \BibitemOpen
  \bibfield  {author} {\bibinfo {author} {\bibfnamefont {T.}~\bibnamefont
  {Mori}}\ and\ \bibinfo {author} {\bibfnamefont {T.}~\bibnamefont {Shirai}},\
  }\href {https://link.aps.org/doi/10.1103/PhysRevLett.125.230604} {\bibfield
  {journal} {\bibinfo  {journal} {Phys. Rev. Lett.}\ }\textbf {\bibinfo
  {volume} {125}},\ \bibinfo {pages} {230604} (\bibinfo {year}
  {2020})}\BibitemShut {NoStop}%
\bibitem [{\citenamefont {S{\'a}}\ \emph {et~al.}(2020)\citenamefont {S{\'a}},
  \citenamefont {Ribeiro},\ and\ \citenamefont {Prosen}}]{sa2020complex}%
  \BibitemOpen
  \bibfield  {author} {\bibinfo {author} {\bibfnamefont {L.}~\bibnamefont
  {S{\'a}}}, \bibinfo {author} {\bibfnamefont {P.}~\bibnamefont {Ribeiro}},\
  and\ \bibinfo {author} {\bibfnamefont {T.}~\bibnamefont {Prosen}},\ }\href
  {https://link.aps.org/doi/10.1103/PhysRevX.10.021019} {\bibfield  {journal}
  {\bibinfo  {journal} {Phys. Rev. X}\ }\textbf {\bibinfo {volume} {10}},\
  \bibinfo {pages} {021019} (\bibinfo {year} {2020})}\BibitemShut {NoStop}%
\bibitem [{\citenamefont {Haga}\ \emph {et~al.}(2021)\citenamefont {Haga},
  \citenamefont {Nakagawa}, \citenamefont {Hamazaki},\ and\ \citenamefont
  {Ueda}}]{haga2021liouvillian}%
  \BibitemOpen
  \bibfield  {author} {\bibinfo {author} {\bibfnamefont {T.}~\bibnamefont
  {Haga}}, \bibinfo {author} {\bibfnamefont {M.}~\bibnamefont {Nakagawa}},
  \bibinfo {author} {\bibfnamefont {R.}~\bibnamefont {Hamazaki}},\ and\
  \bibinfo {author} {\bibfnamefont {M.}~\bibnamefont {Ueda}},\ }\href
  {https://link.aps.org/doi/10.1103/PhysRevLett.127.070402} {\bibfield
  {journal} {\bibinfo  {journal} {Phys. Rev. Lett.}\ }\textbf {\bibinfo
  {volume} {127}},\ \bibinfo {pages} {070402} (\bibinfo {year}
  {2021})}\BibitemShut {NoStop}%
\bibitem [{\citenamefont {Nakagawa}\ \emph {et~al.}(2021)\citenamefont
  {Nakagawa}, \citenamefont {Kawakami},\ and\ \citenamefont
  {Ueda}}]{nakagawa2021exact}%
  \BibitemOpen
  \bibfield  {author} {\bibinfo {author} {\bibfnamefont {M.}~\bibnamefont
  {Nakagawa}}, \bibinfo {author} {\bibfnamefont {N.}~\bibnamefont {Kawakami}},\
  and\ \bibinfo {author} {\bibfnamefont {M.}~\bibnamefont {Ueda}},\ }\href
  {https://link.aps.org/doi/10.1103/PhysRevLett.126.110404} {\bibfield
  {journal} {\bibinfo  {journal} {Phys. Rev. Lett.}\ }\textbf {\bibinfo
  {volume} {126}},\ \bibinfo {pages} {110404} (\bibinfo {year}
  {2021})}\BibitemShut {NoStop}%
\bibitem [{\citenamefont {Diehl}\ \emph {et~al.}(2011)\citenamefont {Diehl},
  \citenamefont {Rico}, \citenamefont {Baranov},\ and\ \citenamefont
  {Zoller}}]{diehl2011topology}%
  \BibitemOpen
  \bibfield  {author} {\bibinfo {author} {\bibfnamefont {S.}~\bibnamefont
  {Diehl}}, \bibinfo {author} {\bibfnamefont {E.}~\bibnamefont {Rico}},
  \bibinfo {author} {\bibfnamefont {M.~A.}\ \bibnamefont {Baranov}},\ and\
  \bibinfo {author} {\bibfnamefont {P.}~\bibnamefont {Zoller}},\ }\href
  {https://doi.org/10.1038/nphys2106} {\bibfield  {journal} {\bibinfo
  {journal} {Nat, Phys.}\ }\textbf {\bibinfo {volume} {7}},\ \bibinfo {pages}
  {971} (\bibinfo {year} {2011})}\BibitemShut {NoStop}%
\bibitem [{\citenamefont {Bardyn}\ \emph {et~al.}(2013)\citenamefont {Bardyn},
  \citenamefont {Baranov}, \citenamefont {Kraus}, \citenamefont {Rico},
  \citenamefont {{\.I}mamo{\u{g}}lu}, \citenamefont {Zoller},\ and\
  \citenamefont {Diehl}}]{bardyn2013topology}%
  \BibitemOpen
  \bibfield  {author} {\bibinfo {author} {\bibfnamefont {C.-E.}\ \bibnamefont
  {Bardyn}}, \bibinfo {author} {\bibfnamefont {M.~A.}\ \bibnamefont {Baranov}},
  \bibinfo {author} {\bibfnamefont {C.~V.}\ \bibnamefont {Kraus}}, \bibinfo
  {author} {\bibfnamefont {E.}~\bibnamefont {Rico}}, \bibinfo {author}
  {\bibfnamefont {A.}~\bibnamefont {{\.I}mamo{\u{g}}lu}}, \bibinfo {author}
  {\bibfnamefont {P.}~\bibnamefont {Zoller}},\ and\ \bibinfo {author}
  {\bibfnamefont {S.}~\bibnamefont {Diehl}},\ }\href
  {https://doi.org/10.1088/1367-2630/15/8/085001} {\bibfield  {journal}
  {\bibinfo  {journal} {New J. Phys.}\ }\textbf {\bibinfo {volume} {15}},\
  \bibinfo {pages} {085001} (\bibinfo {year} {2013})}\BibitemShut {NoStop}%
\bibitem [{\citenamefont {Rivas}\ \emph {et~al.}(2013)\citenamefont {Rivas},
  \citenamefont {Viyuela},\ and\ \citenamefont
  {Martin-Delgado}}]{rivas2013density}%
  \BibitemOpen
  \bibfield  {author} {\bibinfo {author} {\bibfnamefont {A.}~\bibnamefont
  {Rivas}}, \bibinfo {author} {\bibfnamefont {O.}~\bibnamefont {Viyuela}},\
  and\ \bibinfo {author} {\bibfnamefont {M.~A.}\ \bibnamefont
  {Martin-Delgado}},\ }\href
  {https://link.aps.org/doi/10.1103/PhysRevB.88.155141} {\bibfield  {journal}
  {\bibinfo  {journal} {Physical review B}\ }\textbf {\bibinfo {volume} {88}},\
  \bibinfo {pages} {155141} (\bibinfo {year} {2013})}\BibitemShut {NoStop}%
\bibitem [{\citenamefont {Budich}\ \emph {et~al.}(2015)\citenamefont {Budich},
  \citenamefont {Zoller},\ and\ \citenamefont {Diehl}}]{budich2015dissipative}%
  \BibitemOpen
  \bibfield  {author} {\bibinfo {author} {\bibfnamefont {J.~C.}\ \bibnamefont
  {Budich}}, \bibinfo {author} {\bibfnamefont {P.}~\bibnamefont {Zoller}},\
  and\ \bibinfo {author} {\bibfnamefont {S.}~\bibnamefont {Diehl}},\ }\href
  {https://doi.org/10.1103/PhysRevA.91.042117} {\bibfield  {journal} {\bibinfo
  {journal} {Phys. Rev. A}\ }\textbf {\bibinfo {volume} {91}},\ \bibinfo
  {pages} {042117} (\bibinfo {year} {2015})}\BibitemShut {NoStop}%
\bibitem [{\citenamefont {Zhang}\ and\ \citenamefont
  {Gong}(2018)}]{zhang2018topological}%
  \BibitemOpen
  \bibfield  {author} {\bibinfo {author} {\bibfnamefont {D.-J.}\ \bibnamefont
  {Zhang}}\ and\ \bibinfo {author} {\bibfnamefont {J.}~\bibnamefont {Gong}},\
  }\href {https://doi.org/10.1103/PhysRevA.98.052101} {\bibfield  {journal}
  {\bibinfo  {journal} {Phys. Rev. A}\ }\textbf {\bibinfo {volume} {98}},\
  \bibinfo {pages} {052101} (\bibinfo {year} {2018})}\BibitemShut {NoStop}%
\bibitem [{\citenamefont {Altland}\ \emph {et~al.}(2021)\citenamefont
  {Altland}, \citenamefont {Fleischhauer},\ and\ \citenamefont
  {Diehl}}]{altland2021symmetry}%
  \BibitemOpen
  \bibfield  {author} {\bibinfo {author} {\bibfnamefont {A.}~\bibnamefont
  {Altland}}, \bibinfo {author} {\bibfnamefont {M.}~\bibnamefont
  {Fleischhauer}},\ and\ \bibinfo {author} {\bibfnamefont {S.}~\bibnamefont
  {Diehl}},\ }\href {https://link.aps.org/doi/10.1103/PhysRevX.11.021037}
  {\bibfield  {journal} {\bibinfo  {journal} {Phys. Rev. X}\ }\textbf {\bibinfo
  {volume} {11}},\ \bibinfo {pages} {021037} (\bibinfo {year}
  {2021})}\BibitemShut {NoStop}%
\bibitem [{\citenamefont {Dangel}\ \emph {et~al.}(2018)\citenamefont {Dangel},
  \citenamefont {Wagner}, \citenamefont {Cartarius}, \citenamefont {Main},\
  and\ \citenamefont {Wunner}}]{dangel2018topological}%
  \BibitemOpen
  \bibfield  {author} {\bibinfo {author} {\bibfnamefont {F.}~\bibnamefont
  {Dangel}}, \bibinfo {author} {\bibfnamefont {M.}~\bibnamefont {Wagner}},
  \bibinfo {author} {\bibfnamefont {H.}~\bibnamefont {Cartarius}}, \bibinfo
  {author} {\bibfnamefont {J.}~\bibnamefont {Main}},\ and\ \bibinfo {author}
  {\bibfnamefont {G.}~\bibnamefont {Wunner}},\ }\href
  {https://link.aps.org/doi/10.1103/PhysRevA.98.013628} {\bibfield  {journal}
  {\bibinfo  {journal} {Phys. Rev. A}\ }\textbf {\bibinfo {volume} {98}},\
  \bibinfo {pages} {013628} (\bibinfo {year} {2018})}\BibitemShut {NoStop}%
\bibitem [{\citenamefont {van Caspel}\ \emph {et~al.}(2019)\citenamefont {van
  Caspel}, \citenamefont {Arze},\ and\ \citenamefont
  {Castillo}}]{van2019dynamical}%
  \BibitemOpen
  \bibfield  {author} {\bibinfo {author} {\bibfnamefont {M.}~\bibnamefont {van
  Caspel}}, \bibinfo {author} {\bibfnamefont {S.~E.~T.}\ \bibnamefont {Arze}},\
  and\ \bibinfo {author} {\bibfnamefont {I.~P.}\ \bibnamefont {Castillo}},\
  }\href {https://doi.org/10.21468/SciPostPhys.6.2.026} {\bibfield  {journal}
  {\bibinfo  {journal} {SciPost Phys.}\ }\textbf {\bibinfo {volume} {6}},\
  \bibinfo {pages} {26} (\bibinfo {year} {2019})}\BibitemShut {NoStop}%
\bibitem [{\citenamefont {Song}\ \emph {et~al.}(2019)\citenamefont {Song},
  \citenamefont {Yao},\ and\ \citenamefont {Wang}}]{song2019non}%
  \BibitemOpen
  \bibfield  {author} {\bibinfo {author} {\bibfnamefont {F.}~\bibnamefont
  {Song}}, \bibinfo {author} {\bibfnamefont {S.}~\bibnamefont {Yao}},\ and\
  \bibinfo {author} {\bibfnamefont {Z.}~\bibnamefont {Wang}},\ }\href
  {https://link.aps.org/doi/10.1103/PhysRevLett.123.170401} {\bibfield
  {journal} {\bibinfo  {journal} {Phys. Rev. Lett.}\ }\textbf {\bibinfo
  {volume} {123}},\ \bibinfo {pages} {170401} (\bibinfo {year}
  {2019})}\BibitemShut {NoStop}%
\bibitem [{\citenamefont {Goldstein}(2019)}]{goldstein2019dissipation}%
  \BibitemOpen
  \bibfield  {author} {\bibinfo {author} {\bibfnamefont {M.}~\bibnamefont
  {Goldstein}},\ }\href {http://dx.doi.org/10.21468/SciPostPhys.7.5.067}
  {\bibfield  {journal} {\bibinfo  {journal} {SciPost Phys.}\ }\textbf
  {\bibinfo {volume} {7}},\ \bibinfo {pages} {067} (\bibinfo {year}
  {2019})}\BibitemShut {NoStop}%
\bibitem [{\citenamefont {Lieu}\ \emph
  {et~al.}(2020{\natexlab{a}})\citenamefont {Lieu}, \citenamefont {McGinley},\
  and\ \citenamefont {Cooper}}]{lieu2020tenfold}%
  \BibitemOpen
  \bibfield  {author} {\bibinfo {author} {\bibfnamefont {S.}~\bibnamefont
  {Lieu}}, \bibinfo {author} {\bibfnamefont {M.}~\bibnamefont {McGinley}},\
  and\ \bibinfo {author} {\bibfnamefont {N.~R.}\ \bibnamefont {Cooper}},\
  }\href {https://link.aps.org/doi/10.1103/PhysRevLett.124.040401} {\bibfield
  {journal} {\bibinfo  {journal} {Phys. Rev. Lett.}\ }\textbf {\bibinfo
  {volume} {124}},\ \bibinfo {pages} {040401} (\bibinfo {year}
  {2020}{\natexlab{a}})}\BibitemShut {NoStop}%
\bibitem [{\citenamefont {Yoshida}\ \emph {et~al.}(2020)\citenamefont
  {Yoshida}, \citenamefont {Kudo}, \citenamefont {Katsura},\ and\ \citenamefont
  {Hatsugai}}]{yoshida2020fate}%
  \BibitemOpen
  \bibfield  {author} {\bibinfo {author} {\bibfnamefont {T.}~\bibnamefont
  {Yoshida}}, \bibinfo {author} {\bibfnamefont {K.}~\bibnamefont {Kudo}},
  \bibinfo {author} {\bibfnamefont {H.}~\bibnamefont {Katsura}},\ and\ \bibinfo
  {author} {\bibfnamefont {Y.}~\bibnamefont {Hatsugai}},\ }\href
  {https://link.aps.org/doi/10.1103/PhysRevResearch.2.033428} {\bibfield
  {journal} {\bibinfo  {journal} {Phys. Rev. Research}\ }\textbf {\bibinfo
  {volume} {2}},\ \bibinfo {pages} {033428} (\bibinfo {year}
  {2020})}\BibitemShut {NoStop}%
\bibitem [{\citenamefont {Huang}\ \emph {et~al.}(2020)\citenamefont {Huang},
  \citenamefont {Yang},\ and\ \citenamefont {Zhang}}]{huang2020quantum}%
  \BibitemOpen
  \bibfield  {author} {\bibinfo {author} {\bibfnamefont {Y.-W.}\ \bibnamefont
  {Huang}}, \bibinfo {author} {\bibfnamefont {P.-Y.}\ \bibnamefont {Yang}},\
  and\ \bibinfo {author} {\bibfnamefont {W.-M.}\ \bibnamefont {Zhang}},\ }\href
  {https://link.aps.org/doi/10.1103/PhysRevB.102.165116} {\bibfield  {journal}
  {\bibinfo  {journal} {Phys. Rev. B}\ }\textbf {\bibinfo {volume} {102}},\
  \bibinfo {pages} {165116} (\bibinfo {year} {2020})}\BibitemShut {NoStop}%
\bibitem [{\citenamefont {Longhi}(2020)}]{longhi2020unraveling}%
  \BibitemOpen
  \bibfield  {author} {\bibinfo {author} {\bibfnamefont {S.}~\bibnamefont
  {Longhi}},\ }\href {https://link.aps.org/doi/10.1103/PhysRevB.102.201103}
  {\bibfield  {journal} {\bibinfo  {journal} {Phys. Rev. B}\ }\textbf {\bibinfo
  {volume} {102}},\ \bibinfo {pages} {201103(R)} (\bibinfo {year}
  {2020})}\BibitemShut {NoStop}%
\bibitem [{\citenamefont {Pan}\ \emph {et~al.}(2021)\citenamefont {Pan},
  \citenamefont {Li},\ and\ \citenamefont {Gong}}]{pan2021point}%
  \BibitemOpen
  \bibfield  {author} {\bibinfo {author} {\bibfnamefont {J.-S.}\ \bibnamefont
  {Pan}}, \bibinfo {author} {\bibfnamefont {L.}~\bibnamefont {Li}},\ and\
  \bibinfo {author} {\bibfnamefont {J.}~\bibnamefont {Gong}},\ }\href
  {https://doi.org/10.1103/PhysRevB.103.205425} {\bibfield  {journal} {\bibinfo
   {journal} {Phys. Rev. B}\ }\textbf {\bibinfo {volume} {103}},\ \bibinfo
  {pages} {205425} (\bibinfo {year} {2021})}\BibitemShut {NoStop}%
\bibitem [{\citenamefont {Flynn}\ \emph {et~al.}(2021)\citenamefont {Flynn},
  \citenamefont {Cobanera},\ and\ \citenamefont {Viola}}]{flynn2021topology}%
  \BibitemOpen
  \bibfield  {author} {\bibinfo {author} {\bibfnamefont {V.~P.}\ \bibnamefont
  {Flynn}}, \bibinfo {author} {\bibfnamefont {E.}~\bibnamefont {Cobanera}},\
  and\ \bibinfo {author} {\bibfnamefont {L.}~\bibnamefont {Viola}},\ }\href
  {https://link.aps.org/doi/10.1103/PhysRevLett.127.245701} {\bibfield
  {journal} {\bibinfo  {journal} {Phys. Rev. Lett.}\ }\textbf {\bibinfo
  {volume} {127}},\ \bibinfo {pages} {245701} (\bibinfo {year}
  {2021})}\BibitemShut {NoStop}%
\bibitem [{\citenamefont {Prosen}(2012)}]{prosen2012p}%
  \BibitemOpen
  \bibfield  {author} {\bibinfo {author} {\bibfnamefont {T.}~\bibnamefont
  {Prosen}},\ }\href {https://link.aps.org/doi/10.1103/PhysRevLett.109.090404}
  {\bibfield  {journal} {\bibinfo  {journal} {Phys. Rev. Lett.}\ }\textbf
  {\bibinfo {volume} {109}},\ \bibinfo {pages} {090404} (\bibinfo {year}
  {2012})}\BibitemShut {NoStop}%
\bibitem [{\citenamefont {Bu{\v{c}}a}\ and\ \citenamefont
  {Prosen}(2012)}]{buvca2012note}%
  \BibitemOpen
  \bibfield  {author} {\bibinfo {author} {\bibfnamefont {B.}~\bibnamefont
  {Bu{\v{c}}a}}\ and\ \bibinfo {author} {\bibfnamefont {T.}~\bibnamefont
  {Prosen}},\ }\href {https://doi.org/10.1088/1367-2630/14/7/073007} {\bibfield
   {journal} {\bibinfo  {journal} {New J. Phys.}\ }\textbf {\bibinfo {volume}
  {14}},\ \bibinfo {pages} {073007} (\bibinfo {year} {2012})}\BibitemShut
  {NoStop}%
\bibitem [{\citenamefont {Albert}\ and\ \citenamefont
  {Jiang}(2014)}]{albert2014symmetries}%
  \BibitemOpen
  \bibfield  {author} {\bibinfo {author} {\bibfnamefont {V.~V.}\ \bibnamefont
  {Albert}}\ and\ \bibinfo {author} {\bibfnamefont {L.}~\bibnamefont {Jiang}},\
  }\href {https://doi.org/10.1103/PhysRevA.89.022118} {\bibfield  {journal}
  {\bibinfo  {journal} {Phys. Rev. A}\ }\textbf {\bibinfo {volume} {89}},\
  \bibinfo {pages} {022118} (\bibinfo {year} {2014})}\BibitemShut {NoStop}%
\bibitem [{\citenamefont {van Caspel}\ and\ \citenamefont
  {Gritsev}(2018)}]{van2018symmetry}%
  \BibitemOpen
  \bibfield  {author} {\bibinfo {author} {\bibfnamefont {M.}~\bibnamefont {van
  Caspel}}\ and\ \bibinfo {author} {\bibfnamefont {V.}~\bibnamefont
  {Gritsev}},\ }\href {https://link.aps.org/doi/10.1103/PhysRevA.97.052106}
  {\bibfield  {journal} {\bibinfo  {journal} {Phys. Rev. A}\ }\textbf {\bibinfo
  {volume} {97}},\ \bibinfo {pages} {052106} (\bibinfo {year}
  {2018})}\BibitemShut {NoStop}%
\bibitem [{\citenamefont {Huber}\ \emph {et~al.}(2020)\citenamefont {Huber},
  \citenamefont {Kirton}, \citenamefont {Rotter},\ and\ \citenamefont
  {Rabl}}]{10.21468/SciPostPhys.9.4.052}%
  \BibitemOpen
  \bibfield  {author} {\bibinfo {author} {\bibfnamefont {J.}~\bibnamefont
  {Huber}}, \bibinfo {author} {\bibfnamefont {P.}~\bibnamefont {Kirton}},
  \bibinfo {author} {\bibfnamefont {S.}~\bibnamefont {Rotter}},\ and\ \bibinfo
  {author} {\bibfnamefont {P.}~\bibnamefont {Rabl}},\ }\href
  {https://doi.org/10.21468/SciPostPhys.9.4.052} {\bibfield  {journal}
  {\bibinfo  {journal} {SciPost Phys.}\ }\textbf {\bibinfo {volume} {9}},\
  \bibinfo {pages} {52} (\bibinfo {year} {2020})}\BibitemShut {NoStop}%
\bibitem [{\citenamefont {Lieu}\ \emph
  {et~al.}(2020{\natexlab{b}})\citenamefont {Lieu}, \citenamefont {Belyansky},
  \citenamefont {Young}, \citenamefont {Lundgren}, \citenamefont {Albert},\
  and\ \citenamefont {Gorshkov}}]{lieu2020symmetry}%
  \BibitemOpen
  \bibfield  {author} {\bibinfo {author} {\bibfnamefont {S.}~\bibnamefont
  {Lieu}}, \bibinfo {author} {\bibfnamefont {R.}~\bibnamefont {Belyansky}},
  \bibinfo {author} {\bibfnamefont {J.~T.}\ \bibnamefont {Young}}, \bibinfo
  {author} {\bibfnamefont {R.}~\bibnamefont {Lundgren}}, \bibinfo {author}
  {\bibfnamefont {V.~V.}\ \bibnamefont {Albert}},\ and\ \bibinfo {author}
  {\bibfnamefont {A.~V.}\ \bibnamefont {Gorshkov}},\ }\href
  {https://link.aps.org/doi/10.1103/PhysRevLett.125.240405} {\bibfield
  {journal} {\bibinfo  {journal} {Phys. Rev. Lett.}\ }\textbf {\bibinfo
  {volume} {125}},\ \bibinfo {pages} {240405} (\bibinfo {year}
  {2020}{\natexlab{b}})}\BibitemShut {NoStop}%
\bibitem [{pc()}]{pc}%
  \BibitemOpen
  \href@noop {} {}\bibinfo {note} {{There is an additional physical condition
  for the Lindbladian spectra. Since the solution of the GKSL equation to be
  Hermitian, the eigenvalues of general Lindbladians must form anti-conjugation
  pairs \((\nu,-\nu^*)\).}}\BibitemShut {Stop}%
\bibitem [{\citenamefont {Prosen}(2008)}]{prosen2008third}%
  \BibitemOpen
  \bibfield  {author} {\bibinfo {author} {\bibfnamefont {T.}~\bibnamefont
  {Prosen}},\ }\href {https://doi.org/10.1088/1367-2630/10/4/043026} {\bibfield
   {journal} {\bibinfo  {journal} {New J. Phys.}\ }\textbf {\bibinfo {volume}
  {10}},\ \bibinfo {pages} {043026} (\bibinfo {year} {2008})}\BibitemShut
  {NoStop}%
\bibitem [{\citenamefont {Prosen}(2010)}]{prosen2010spectral}%
  \BibitemOpen
  \bibfield  {author} {\bibinfo {author} {\bibfnamefont {T.}~\bibnamefont
  {Prosen}},\ }\href {https://doi.org/10.1088/1742-5468/2010/07/P07020}
  {\bibfield  {journal} {\bibinfo  {journal} {J. Stat. Mech.}\ }\textbf
  {\bibinfo {volume} {2010}},\ \bibinfo {pages} {P07020} (\bibinfo {year}
  {2010})}\BibitemShut {NoStop}%
\bibitem [{\citenamefont {Su}\ \emph {et~al.}(1980)\citenamefont {Su},
  \citenamefont {Schrieffer},\ and\ \citenamefont {Heeger}}]{su1980soliton}%
  \BibitemOpen
  \bibfield  {author} {\bibinfo {author} {\bibfnamefont {W.-P.}\ \bibnamefont
  {Su}}, \bibinfo {author} {\bibfnamefont {J.~R.}\ \bibnamefont {Schrieffer}},\
  and\ \bibinfo {author} {\bibfnamefont {A.~J.}\ \bibnamefont {Heeger}},\
  }\href {https://link.aps.org/doi/10.1103/PhysRevB.22.2099} {\bibfield
  {journal} {\bibinfo  {journal} {Phys. Rev. B}\ }\textbf {\bibinfo {volume}
  {22}},\ \bibinfo {pages} {2099} (\bibinfo {year} {1980})}\BibitemShut
  {NoStop}%
\bibitem [{\citenamefont {Kitaev}(2001)}]{kitaev2001unpaired}%
  \BibitemOpen
  \bibfield  {author} {\bibinfo {author} {\bibfnamefont {A.~Y.}\ \bibnamefont
  {Kitaev}},\ }\href {https://doi.org/10.1070/1063-7869/44/10S/S29} {\bibfield
  {journal} {\bibinfo  {journal} {Phys. Usp.}\ }\textbf {\bibinfo {volume}
  {44}},\ \bibinfo {pages} {131} (\bibinfo {year} {2001})}\BibitemShut
  {NoStop}%
\bibitem [{sta()}]{state}%
  \BibitemOpen
  \href@noop {} {}\bibinfo {note} {{We note that the eigenvectors of \(Z\) are
  not proper quantum states in general because \(\rho_j\) in Eq.\
  \eqref{eq:eigen_lindblad} are not Hermitian when \(\mathrm{Re}\,\nu_j\ne0\).
  However, we call the eigenvectors as "states" to adapt to the convention of
  topological matters.}}\BibitemShut {Stop}%
\end{thebibliography}%

\end{document}